\documentclass[twocolumn]{aastex631}

\shorttitle{Massive molecular gas in central galaxies}
\shortauthors{Fujita et al.}

 \usepackage[T1]{fontenc}

\begin{document}

\title{Massive Molecular Gas as a Fuel Tank for Active Galactic Nuclei Feedback in
Central Cluster Galaxies}

\author[0000-0003-0058-9719]{Yutaka Fujita}
\affiliation{Department of Physics, Graduate School of Science, 
Tokyo Metropolitan University,\\
1-1 Minami-Osawa, Hachioji-shi, Tokyo 192-0397, Japan}

\author[0000-0003-2535-5513]{Nozomu Kawakatu}
\affiliation{National Institute of Technology, Kure College, \\
2-2-11,
Agaminami, Kure, Hiroshima, 737-8506, Japan}

\author[0000-0003-0292-3645]{Hiroshi Nagai} \affiliation{National
Astronomical Observatory of Japan, Osawa 2-21-1, Mitaka, Tokyo 181-8588,
Japan} \affiliation{The Graduate University for Advanced Studies,
SOKENDAI, Osawa 2-21-1, Mitaka, Tokyo 181-8588, Japan}

\begin{abstract}
Massive molecular gas has been discovered in giant elliptical galaxies
at the centers of galaxy clusters. To reveal its role in active galactic
nucleus (AGN) feedback in those galaxies, we construct a semianalytical
model of gas circulation. This model especially focuses on the massive
molecular gas (interstellar cold gas on a scale of $\sim 10$~kpc) and
the circumnuclear disk ($\lesssim 0.5$~kpc). We consider the destruction
of the interstellar cold gas by star formation and the gravitational
instability for the circumnuclear disk. Our model can reproduce the
basic properties of the interstellar cold gas and the circumnuclear
disk, such as their masses. We also find that the circumnuclear disk
tends to stay at the boundary between stable and unstable states. This
works as an 'adjusting valve' that regulates mass accretion toward the
supermassive black hole. On the other hand, the interstellar cold gas
serves as a 'fuel tank' in the AGN feedback. Even if the cooling of the
galactic hot gas is prevented, the interstellar cold gas can sustain the
AGN activity for $\gtrsim 0.5$~Gyr. We also confirm that the small
entropy of the hot gas ($\lesssim 30\rm\: keV\: cm^2$) or the short
cooling time ($\lesssim 1$~Gyr) is a critical condition for the
existence of the massive amounts of molecular gas in the galaxy. The
dissipation time of the interstellar cold gas may be related to the
critical cooling time. The galaxy behavior is described by a simple
relation among the disk stability, the cloud dissipation time, and the
gas cooling rate.
\end{abstract}


\section{Introduction} 
\label{sec:intro}

Radiative cooling time of the hot intracluster medium (ICM) is often
shorter than the Hubble time in the cores of many galaxy clusters. In
the absence of any heating sources, the hot gas in the core should cool,
and a 'cooling flow' toward the cluster center should develop
\citep{1994ARA&A..32..277F}.  However, X-ray observations have denied
the existence of massive cooling flows in clusters, indicating that the
cores are heated by some unknown source
\citep[e.g.][]{1997ApJ...481..660I,2001A&A...365L..99K,2001A&A...365L.104P,2001A&A...365L..87T}. The
active galactic nucleus (AGN) that resides in the central galaxy of a
cluster is the most promising candidate of the heating source
\citep[AGN feedback;][]{2000A&A...356..788C,2007ARA&A..45..117M,2012ARA&A..50..455F}.

In addition to the hot X-ray gas, much cooler gas has been discovered in
the central galaxies in the core of galaxy clusters. For example,
massive molecular clouds ($\gtrsim 10^9\: M_\odot$) have been detected,
although the mass is much smaller than the prediction of a classical
cooling flow model
\citep{2001MNRAS.328..762E,2003A&A...412..657S,2014ApJ...792...94D,2014ApJ...785...44M,2014ApJ...784...78R,2016MNRAS.458.3134R,2016Natur.534..218T,2016ApJ...832..148V}. Nebular emission
associated with warm gas has also been observed
\citep{1989ApJ...338...48H,1999MNRAS.306..857C,2010ApJ...721.1262M,2015MNRAS.451.3768T}.
It has been indicated that the nebular emission, enhanced star formation,
and AGN activity are tend to be observed in cluster cores when the
central entropy drops down to $\lesssim 30 \rm\: keV\: cm^2$, or almost
equivalently when the central cooling time is $\lesssim 1$~Gyr
\citep{2008ApJ...683L.107C,2008ApJ...687..899R,2009MNRAS.398.1698S,2017MNRAS.464.4360M}.

While some of the molecular gas in the central galaxy is consumed in
star formation, part of it may accrete onto the supermassive black hole
(SMBH) at the galactic center. In fact, it has been discussed that the
cold gas, rather than the hot gas, fuels the SMBH
\citep{1989Natur.338...45S,2010MNRAS.408..961P}.  However, the angular
momentum would hinder the radial infall of the gas, which leads to the
formation of a circumnuclear disk around the SMBH ($\lesssim
0.5$~kpc). Active star formation has been observed in the circumnuclear
disk as nuclear starbursts in nearby galaxies
\citep{2004ApJ...617..214I,2007ApJ...671.1388D,2008ApJ...677..895W}.
Previous studies have indicated that the star formation can induce the
mass accretion onto the SMBH
\citep[e.g.][]{1997ApJ...479L..97U,2002ApJ...566L..21W,2005ApJ...630..167T,2007MNRAS.374..515L,2008A&A...491..441V,2010MNRAS.404.2170K}. In
particular, the turbulence generated by supernova (SN) explosions
associated with the star formation may regulate the kinetic viscosity of
the disk and may change the mass accretion rate
\citep{2008ApJ...681...73K,2013A&A...560A..34W}. This means that the
tiny circumnuclear disk may control AGN feedback and may affect the
evolution of the central galaxy and the cluster core.

In this paper, we study AGN feedback and gas circulation of the central
galaxies in cluster cores using a semianalytical model. We employ a
semianalytical model because although a number of numerical simulations
have been performed for the formation of cold gas and the AGN feedback
\citep[e.g.][]{2007ApJ...665.1038C,2010ApJ...717..708C,2012MNRAS.424..728B,2012MNRAS.419.3319M,2012MNRAS.420.3174S,2013MNRAS.432.3401G,2013MNRAS.430..174H,2014ApJ...780..126G,2015ApJ...808...43M,2017MNRAS.472.4707B,2017ApJ...835...15C,2017MNRAS.470.4530W,2019MNRAS.490..343B,2019ApJ...872L..11Q,2019ApJ...877...47Q,2020NatAs...4..900Q,2021MNRAS.501..398Y},
it is still difficult to cover the whole scale from the cluster core
($\sim 30$~kpc) to the star formation and turbulence in the
circumnuclear disk ($\lesssim$~pc).  Moreover, it is much easier to
interpret the results of semianalytical models than those of numerical
simulations. In this study, we focus on the roles of the massive
molecular gas and the circumnuclear disk in the AGN feedback. We
construct our model based on representative models of molecular gas
evolution \citep{1997ApJ...480..235E} and the circumnuclear disk
\citep{2008ApJ...681...73K,2013A&A...560A..34W}. This paper is organized
as follows. In Section \ref{sec:model}, we describe our models for the
gas circulation and the AGN feedback.  In
Section~\ref{sec:behav}, we indicate that the resulting behavior of the
model galaxy can be represented by a simple relation. In
Section~\ref{sec:result}, we show the results of our fiducial model. In
Section~\ref{sec:discuss}, we discuss the roles of the massive molecular
gas clouds and the circumnuclear disk in the AGN feedback as well as the
influence of the entropy of the hot gas. The conclusion of this paper is
presented in Section~\ref{sec:conc}.

\section{Models}
\label{sec:model}

\subsection{Overview}

\begin{figure}
\includegraphics[width=84mm]{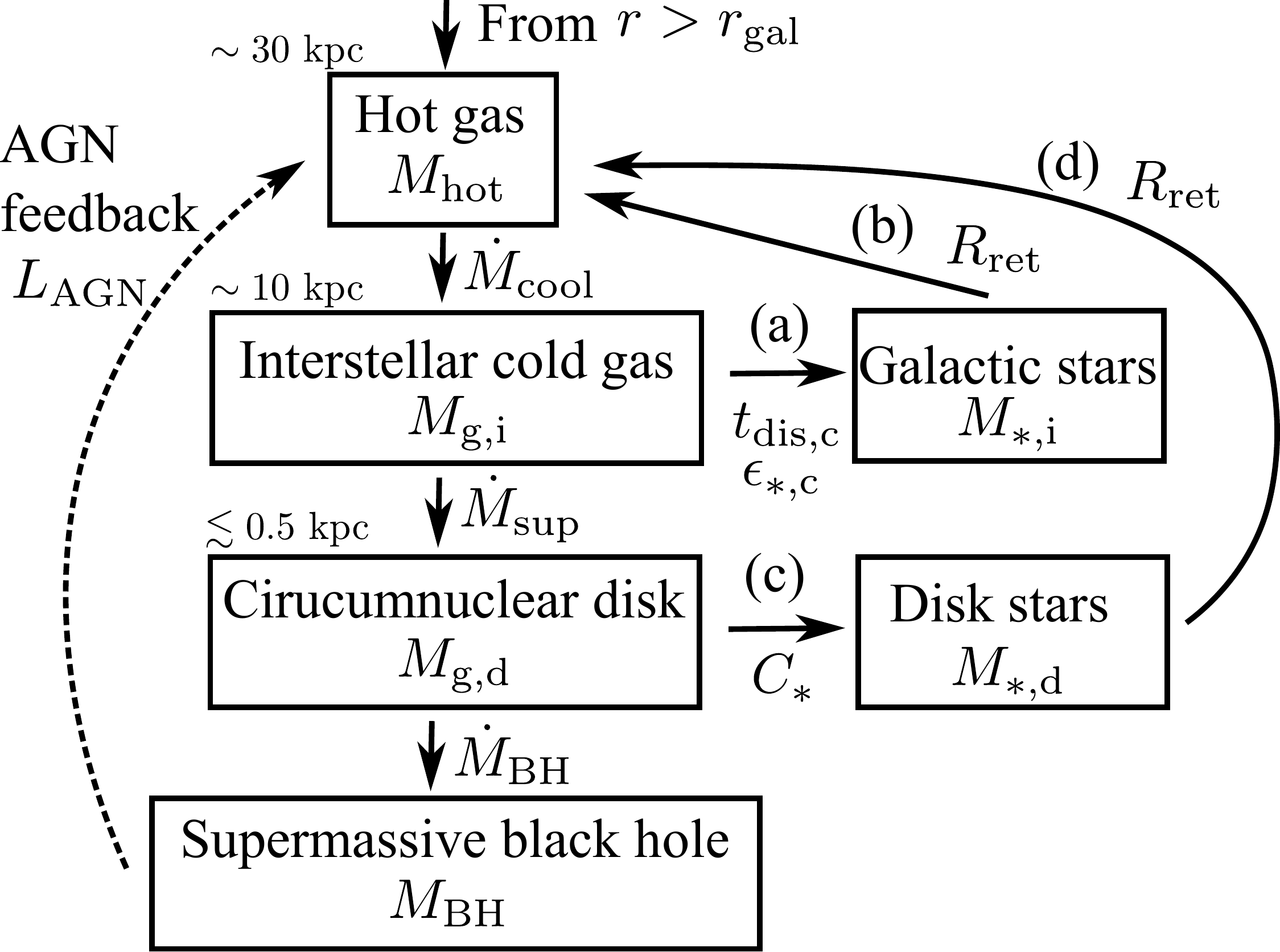} \caption{Flow chart showing the
different components in the galaxy. The solid arrows show gas flows, and
the dotted arrow shows the AGN feedback. The spacial scales of the hot gas, the
interstellar cold gas, and the circumnuclear disk are presented.}
\label{fig:flow}
\end{figure}

Figure~\ref{fig:flow} shows a flow chart of the relations among
different gas components. Although we later construct evolution
equations following this chart (see Section~\ref{eq:evo}), here we
briefly explain the whole picture of gas flows among the components. The
'hot gas' in the galaxy cools through radiative cooling and becomes the
'interstellar cold gas', which corresponds to the observed massive
molecular gas ($\gtrsim 10^9\: M_\odot$ see
Section~\ref{sec:intro}). Thus, the mass of the interstellar cold gas
($M_{\rm g,i}$) depends on the supply from the hot gas ($\dot{M}_{\rm
cool}$). While most of the interstellar cold gas turns into galactic
stars with a disruption time-scale $t_{\rm dis,c}$ and a star formation
efficiency $\epsilon_{\rm *,c}$ (arrow (a) in Figure~\ref{fig:flow}),
the rest eventually flows into the galactic center at a rate of
$\dot{M}_{\rm sup}$ and forms the 'circumnuclear disk' around the
SMBH. The mass of the disk is represented by $M_{\rm g,d}$ and it is
typically $\sim 10^8\: M_\odot$ \citep{2019ApJ...883..193N}. Because the
circumnuclear disk can also contain molecular gas, we use the term
'interstellar cold gas' instead of molecular gas. Some of the disk gas
pours into the SMBH at a rate of $\dot{M}_{\rm BH}$ and the rest is
consumed in star formation in the disk with an efficiency of $C_*$ (arrow
(c) in Figure~\ref{fig:flow}). The gas attracted by the SMBH activates
the AGN, which prevents the galactic hot gas from cooling further
(dashed-arrow in Figure~\ref{fig:flow}). The energy injection rate of
the AGN is represented by $L_{\rm AGN}$. The masses of the galactic and
the disk stars are represented by $M_{\rm *,i}$ and $M_{\rm *,d}$,
respectively. Some fraction ($R_{\rm ret}$) of the stellar mass goes
back to the hot gas (arrows (b) and (d) in Figure~\ref{fig:flow}). In
the following subsections, we describe the details of each
component. The list of main parameters is shown in
Table~\ref{tab:symbol}.

\begin{deluxetable*}{ccc}
\tabletypesize{\scriptsize}
\tablecaption{List of main parameters and symbols\label{tab:symbol}}
\tablewidth{0pt}
\tablehead{\colhead{Parameter} & \colhead{Symbol} & \colhead{Equations or sections}}
\decimalcolnumbers
\startdata
Potential of the galaxy component & $\Phi_{\rm ISO}$ & (\ref{eq:PhiISO}) \\
Potential of the cluster component & $\Phi_{\rm NFW}$ & (\ref{eq:PhiNFW}) \\
Potential of newly formed stars & $\Phi_{\rm nstar}$ & (\ref{eq:Phi}) \\
Total potential & $\Phi$ & (\ref{eq:Phi}) \\
Total mass  & $M_{\rm host}$ & \ref{sec:pot} \\
Galaxy radius & $r_{\rm gal}$ & \ref{sec:hot} \\
Density of the hot gas & $\rho_{\rm hot}$ & (\ref{eq:rhohot}) \\
$\rho_{\rm hot}$ at $r_{\rm gal}$ & $\rho_{\rm hot,0}$ & (\ref{eq:rhohot}) \\
Mass of the hot gas & $M_{\rm hot}$ & (\ref{eq:Mhot}) \\
Virial temperature of the galaxy & $kT_{\rm vir}$ & (\ref{eq:Tvir}) \\
Average hot gas temperature & $kT_{\rm hot}$ & (\ref{eq:Thot}) \\
Cooling function of the hot gas & $\Lambda$ & (\ref{eq:cool}) \\
X-ray luminosity of the galaxy & $L_{\rm cool}$ & (\ref{eq:Lcool}) \\
Effective cooling radius & $r_{\rm cool}$ & (\ref{eq:Lcool}) \\
Electron number density & $n_{\rm e}$ & (\ref{eq:Lcool}) \\
$n_{\rm e}$ at $r_{\rm gal}$ & $n_{\rm e,0}$ & \ref{sec:hot} \\
Cooling rate of the hot gas & $\dot{M}_{\rm cool}$ & (\ref{eq:dMacc}) \\
Unsuppressed $\dot{M}_{\rm cool}$ & $\dot{M}_{\rm cool,0}$ & (\ref{eq:dMacc0}) \\
Mass of a molecular cloud & $M_{\rm c}$ & (\ref{eq:dMc}) \\
Mass of stars in a cloud & $M_{\rm*,c}$ & (\ref{eq:dMc}) \\
Luminosity of stars in a cloud & $L_{\rm *,c}$ & (\ref{eq:dMc}) \\
Turbulent velocity of cloud gas & $v_{\rm t,c}$ & \ref{sec:ICG_evo} \\
Cloud disruption time & $t_{\rm dis,c}$ & (\ref{eq:epssc}) \\
Efficiency of star formation & $\epsilon_{\rm *,c}$ & (\ref{eq:epssc}) \\
Mass of the interstellar cold gas & $M_{\rm g,i}$ & \ref{sec:ICG_evo} \\
Extent of the interstellar cold gas & $r_{\rm i}$ & \ref{sec:coldgas} \\
Cooling time of the  hot gas & $t_{\rm cool}$ & \ref{sec:coldgas} \\
Free-fall time & $t_{\rm ff}$ & \ref{sec:coldgas} \\
Gravity of the galaxy & $g$ & \ref{sec:coldgas} \\
Mass of the SMBH & $M_{\rm BH}$ & (\ref{eq:MBH}) \\
Disk gas mass & $M_{\rm g,d}$ & (\ref{eq:MBH}) \\
Disk radius & $r_{\rm d}$ & (\ref{eq:MBH}) \\
Critical surface density of the disk & $\Sigma_{\rm crit}$ & (\ref{eq:Sigcrit}) \\
Sound velocity of the disk gas & $c_{\rm s,d}$ & (\ref{eq:Sigcrit}) \\
Angular velocity of the disk & $\Omega$ & (\ref{eq:Omega}) \\
Disk gas density & $\rho_{\rm d}$ & (\ref{eq:rhogd}) \\
Turbulent velocity of the disk gas & $v_{\rm t,d}$ & (\ref{eq:rhogd}) \\
Scale height of the disk & $h$ & (\ref{eq:rhogd}),(\ref{eq:rhod}),(\ref{eq:hst}) \\
Surface density of the disk & $\Sigma_{\rm d}$ & (\ref{eq:g}) \\
Vertical gravity of the disk & $g_{\rm d}$ & (\ref{eq:g}) \\
Heating efficiency of SNe & $\eta$  & (\ref{eq:diskbalance1}) \\
Star formation rate of the disk & $S_*$ & (\ref{eq:diskbalance1}) \\
SN energy & $E_{\rm SN}$ & (\ref{eq:diskbalance1}) \\
Dissipation time of disk turbulence & $t_{\rm dis,d}$ & (\ref{eq:diskbalance1}) \\
Star formation efficiency of the disk & $C_*$ & (\ref{eq:cd3}) \\
SMBH accretion rate & $\dot{M}_{\rm BH}$ & (\ref{eq:dotMBH}) \\
Coefficient of disk viscosity & $\nu$ & (\ref{eq:dotMBH}),(\ref{eq:nuv}),(\ref{eq:nuc}) \\
$\alpha$-parameter of the disk & $\alpha$ & (\ref{eq:nuv}),(\ref{eq:nuc}) \\
AGN power & $L_{\rm AGN}$ & (\ref{eq:LAGN}) \\
Heating efficiency of the AGN & $\epsilon_{\rm heat}$ & (\ref{eq:LAGN}) \\
Cooling suppression factor & $f_{\rm sup}$ & (\ref{eq:fsup}) \\
Supply from the cold interstellar gas & $\dot{M}_{\rm sup}$ & (\ref{eq:dMsup}) \\
Return fraction of the stellar mass & $R_{\rm ret}$ & (\ref{eq:dMsi}) \\
Stellar mass in the interstellar cold gas & $M_{\rm *,i}$ & (\ref{eq:dMsi}) \\
Stellar mass in the disk & $M_{\rm *,d}$ & (\ref{eq:dMsd}) \\
Specific entropy & $K$ & \ref{sec:entro} \\
Specific entropy at $r=30$~kpc & $K_{30}$ & \ref{sec:entro} \\
Star formation rate of the galaxy & $ \dot{M}_*$ & (\ref{eq:dMs}) \\
\enddata
\tablecomments{The representative equations or sections in which the parameters appear are shown in the column (3).}
\end{deluxetable*}

\subsection{Galaxy potential}
\label{sec:pot}

We assume that the central galaxy and the host cluster are spherically
symmetric except for the circumnuclear disk, and that the SMBH is
located at $r=0$. The gravitational potential around the galaxy is
mainly composed of two components
\citep{2017ApJ...837...51H,2018ApJ...853..177P}. One is a central-core
isothermal potential representing the galaxy,
\begin{equation}
\label{eq:PhiISO}
 \Phi_{\rm ISO}(r) = -\sigma^2\ln(1 + (r/r_{\rm I})^2)\:,
\end{equation}
where $\sigma$ is the stellar velocity dispersion, and $r_{\rm I}$ is the
core radius. The other is a Navarro–Frenk–White (NFW) potential of the host cluster,
\begin{equation}
\label{eq:PhiNFW}
 \Phi_{\rm NFW}(r) = -4\pi G\rho_0 r_s^2
 \frac{\ln(1+r/r_{\rm s})}{r/r_{\rm s}}\:,
\end{equation}
where $\rho_0$ is the characteristic matter density, and $r_s$ is the
scale radius \citep{1997ApJ...490..493N}. In this paper, we assume that
$\sigma=250\rm\: km\: s^{-1}$, $r_s=400$~kpc, and $4\pi G \rho_0
r_s^2\mu m_p=40$~keV, where $\mu=0.6$ is the mean molecular weight,
and $m_p$ is the proton mass. These values are typical ones for
clusters and the central galaxies
\citep{2017ApJ...837...51H,2018ApJ...853..177P}. For the core radius of
the galaxy, we adopt $r_{\rm I}=0.8$~kpc
\citep[e.g.][]{2005MNRAS.359..755W,2016MNRAS.456.4475O}. We also
consider the gravitational potential associated with newly formed stars
in the interstellar cold gas and the circumnuclear disk, $\Phi_{\rm
nstar}(r)$, which will be given in section~\ref{eq:evo}. The
contribution of gas to the potential is ignored. Thus, the potential of
the whole system is represented by
\begin{equation}
\label{eq:Phi}
 \Phi(r) = \Phi_{\rm ISO}(r) + \Phi_{\rm NFW}(r)
 + \Phi_{\rm nstar}(r)\:,
\end{equation}
and the mass distribution corresponding to $\Phi(r)$ is referred to as
$M_{\rm host}(r)$. 

\subsection{Hot gas}
\label{sec:hot}

We focus on gas circulation and AGN feedback in the innermost region of
a cluster, that is, within the central galaxy. We define the influential
sphere of the central galaxy as $r<r_{\rm gal}$, and we adopt $r_{\rm
gal}=30$~kpc as the 'boundary'. The region $r<r_{\rm gal}$ generally
covers the stellar system of the galaxy
\citep{2016MNRAS.456.4475O}. Since our main interest is the interstellar
cold gas and the circumnuclear disk, we adopt a simplified model for the
hot gas. Thus, we assume that the gas density at the boundary $r=r_{\rm
gal}$ is constant, and the density profile is given by a power law
\begin{equation}
 \label{eq:rhohot}
\rho_{\rm hot}(r)=\rho_{\rm
hot,0}(r/r_{\rm gal})^{-\alpha_{\rm hot}}\:.
\end{equation}
Moreover, in order to avoid complexity, we assume that $\alpha_{\rm
hot}$ is time-independent, which means that the mass of the hot gas for
$r<r_{\rm gal}$ is constant. This implies that, if some of the hot gas
is removed through radiative cooling, the equal amount of hot gas is
supplied from the outside of the boundary ($r>r_{\rm gal}$). While this
may rather be an oversimplification, it allows us to study the influence
of boundary conditions at $r=r_{\rm gal}$ on the results. In our
fiducial model, we assume that the electron density at $r=r_{\rm gal}$
is $n_{\rm e,0}=0.86\: \rho_{\rm hot,0}/m_p = 0.02\rm\: cm^{-3}$ and
that the index is $\alpha_{\rm hot}=1$
(Equation~(\ref{eq:rhohot})). Those are typical values for cool-core
clusters \citep{2017ApJ...837...51H}. The mass of the hot gas is
\begin{equation}
 \label{eq:Mhot}
M_{\rm hot}(r)=4\pi\int_0^r \rho_{\rm hot}(r) r^2 dr\:.
\end{equation}

The temperature of the hot gas basically reflects the virial temperature
of the galaxy, which is given by
\begin{equation}
\label{eq:Tvir}
 kT_{\rm vir} = |\Phi(r_{\rm gal})|/2\:,
\end{equation}
where $k$ is the Boltzmann constant.
Observations have shown that the temperature of the hot gas around the
central galaxies of clusters is a factor of 2 larger than the virial
temperature \citep{2001ApJ...547..693M,2009A&A...501..157N}. Thus, the
hot gas temperature $kT_{\rm hot}$ is given by
\begin{equation}
\label{eq:Thot}
 kT_{\rm hot} = 2\: kT_{\rm vir}
\end{equation}
For the parameters we selected in Section~\ref{sec:pot}, $kT_{\rm
hot}\sim 2.2$~keV.

The hot gas loses its thermal energy through radiative cooling. For the
cooling function, we adopt the following metallicity-dependent one:
\begin{eqnarray}
\Lambda(T,Z)&=&2.41\times 10^{-27}
\left[0.8+0.1\left(\frac{Z}{Z_\odot}\right)\right]
\left(\frac{T}{\rm K}\right)^{0.5}\nonumber\\
&+ & 1.39\times 10^{-16}
\left[0.02+\left(\frac{Z}{Z_\odot}\right)^{0.8}\right]\nonumber\\
& & \times\left(\frac{T}{\rm K}\right)^{-1.0}\rm\: erg\: cm^{3}\:s^{-1}
\label{eq:cool}
\end{eqnarray}
\citep{2013MNRAS.428..599F}, where $T$ is the temperature, and $Z$ is the
metallicity. This function approximates the one derived by
\citet{1993ApJS...88..253S} for $T\ga 10^5$~K and $Z\la 1\:
Z_\odot$. Since we consider the hot gas inside the central galaxy, we
assume that $Z=0.5\: Z_\odot$.

The X-ray luminosity of the galaxy is given by
\begin{equation}
\label{eq:Lcool}
 L_{\rm cool}=\int_0^{r_{\rm cool}}4\pi r^2 n_{\rm e}(r)^2 \Lambda(T_{\rm hot},Z) dr\:,
\end{equation}
where $n_{\rm e}(r)$ is the electron number density, and it is given by
$n_{\rm e}=0.86\:\rho_{\rm hot}/m_p$. The upper limit of the integral
$r_{\rm cool}$ is the radius within which radiative cooling is
effective, which will be specified in Section~\ref{sec:coldgas}. The
cooling rate of the hot gas is written as
\begin{equation}
\label{eq:dMacc}
 \dot{M}_{\rm cool} = \dot{M}_{\rm cool,0} f_{\rm sup}\:,
\end{equation}
where
\begin{equation}
\label{eq:dMacc0}
 \dot{M}_{\rm cool,0} = \frac{2}{5}
\frac{\mu m_p L_{\rm cool}}{kT_{\rm hot}}\:
\end{equation}
\citep[e.g.][]{1994ARA&A..32..277F}, and $f_{\rm sup}$ is the
suppression factor by AGN feedback, which will be given in
Equation~(\ref{eq:fsup}).

\subsection{Interstellar cold gas cloud}
\label{sec:ICG}

\subsubsection{Evolution}
\label{sec:ICG_evo}

The evolution of the interstellar cold gas is affected by the
surrounding pressure and the star formation in the cold gas. We treat
the gas as a massive cold cloud and adopt the model by
\citet{1997ApJ...480..235E}, although it is not obvious whether
this model, which was constructed for Milky Way clouds, can be applied to the
filamentary cold gas observed in cluster cores.

In the model by \citet{1997ApJ...480..235E}, the cloud mass $M_{\rm c}$
evolves as
\begin{equation}
\label{eq:dMc}
 \frac{dM_{\rm c}}{dt_{\rm c}} =
 - \frac{dM_{\rm *,c}}{dt_{\rm c}} - \frac{A L_{\rm *,c}}{v_{\rm t,c}^2}\:,
\end{equation}
where $t_{\rm c}$ is the time lapsed since the cloud is formed, $M_{\rm
*,c}$ is the mass of the stars that are formed in the cloud, $L_{\rm
*,c}$ is the luminosity of the stars, $v_{\rm t,c}$ is the turbulent
velocity of the cloud gas, and $A$ is a dimensionless constant. The
binding energy of the cloud is written as $M_{\rm c}v_{\rm t,c}^2$, and
the turbulent velocity is given by $v_{\rm t,c}\propto (P_{\rm c}M_{\rm
c}^2)^{1/8}$, where $P_{\rm c}$ is the cloud pressure
\citep{1989ApJ...338..178E}. The star formation rate $dM_{\rm
*,c}/dt_{\rm c}$ is assumed to be constant.

\begin{figure}
\includegraphics[width=84mm]{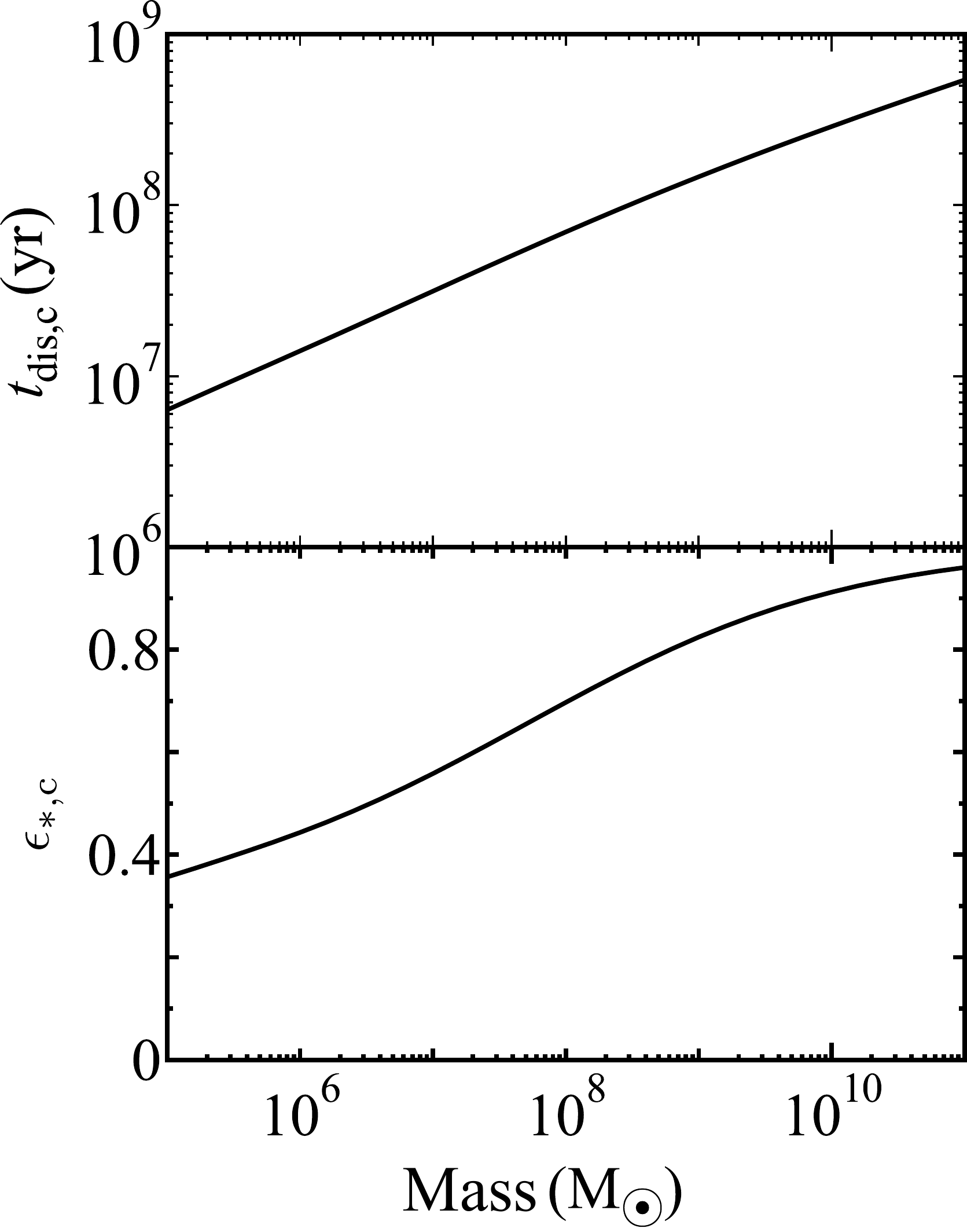} \caption{Disruption time
$t_{\rm dis,c}$ (upper) and $\epsilon_{\rm *,c}$ (lower) represented as
functions of the cloud mass when $P_{\rm c}=100\: P_\odot$.}
\label{fig:tdiscM}
\end{figure}

The cloud disruption time $t_{\rm dis,c}$ is defined as the time when
the cloud mass reaches zero ($M_{\rm c}=0$). The efficiency of star
formation $\epsilon_{\rm *,c}$ is given by
\begin{equation}
\label{eq:epssc}
 \epsilon_{\rm *,c} = \frac{dM_{\rm *,c}}{dt_{\rm c}}\frac{t_{\rm
 dis,c}}{M_{\rm c}(t_{\rm c}=0)}\:.
\end{equation}
Both $t_{\rm dis,c}$ and $\epsilon_{\rm *,c}$ can be derived
analytically as functions of $P_{\rm c}$ and $M_{\rm c}(t_{\rm i}=0)$
and we show the results in Figure~\ref{fig:tdiscM} for
$P_{\rm c}=100\: P_\odot$, where $P_\odot=3\times 10^4\rm\: cm^{-3}\: K$
is the pressure in the solar neighborhood. The disruption time $t_{\rm
dis,c}$ is an increasing function of $M_{\rm c}$ because it is
proportional to the crossing time of a molecular cloud $R_{\rm c}/v_{\rm
t,c}$, where $R_{\rm c}$ is the cloud radius
\citep{1997ApJ...480..235E}. Moreover, the efficiency of star formation
$\epsilon_{\rm *,c}$ is also an increasing function of $M_{\rm c}$
because it depends on the binding energy of the cloud
\citep{1997ApJ...480..235E}.

In this study, we adopt $t_{\rm dis,c}$ and $\epsilon_{\rm *,c}$ for
$P_{\rm c}=100\: P_\odot$ (Figure~\ref{fig:tdiscM}). The adopted value
of $P_{\rm c}$ is roughly consistent with the assumed hot gas pressure
of our model galaxy (Equations~(\ref{eq:rhohot}) and~(\ref{eq:Thot})),
and $t_{\rm dis,c}$ and $\epsilon_{\rm *,c}$ are not much sensitive to
$P_{\rm c}$. Since $P_{\rm c}$ is fixed, the disruption time and the
efficiency can be represented by $t_{\rm dis,c}=t_{\rm dis,c}(M_{\rm
c})$ and $\epsilon_{\rm *,c}=\epsilon_{\rm *,c}(M_{\rm c})$,
respectively. Compared to those for the solar neighborhood
($P=P_\odot$), $t_{\rm dis,c}$ is smaller and $\epsilon_{\rm *,c}$ is
larger at $P_{\rm c}=100\: P_\odot$ for a given $M_{\rm c}$ (see
Figure~4 in \citealt{1997ApJ...480..235E}).

The above model describes the evolution of a cold cloud without an
additional mass supply. For the central galaxy we consider, however, we
need to include the mass supply from hot gas $\dot{M}_{\rm cool}$
(Equation~(\ref{eq:dMacc}) and Figure~\ref{fig:flow}), which means that
the evolution of the cold interstellar gas does not simply follow
Equation~(\ref{eq:dMc}). Thus, when we consider the evolution of the
interstellar cold gas in Section~\ref{eq:evo}, we use the letter $M_{\rm
g,i}$ instead of $M_{\rm c}$ in order to clarify the difference, and we
use the values $t_{\rm dis,c}(M_{\rm g,i})$ and $\epsilon_{\rm
*,c}(M_{\rm g,i})$ as variable parameters in evolution
Equations~(\ref{eq:dMgi})--(\ref{eq:dMsi}). These two
parameters determine the evolution of the interstellar cold gas.

\subsubsection{Spatial distribution}
\label{sec:coldgas}

Observationally, the cold interstellar gas has been detected in the
region where the cooling time of the hot gas is $t_{\rm cool}\lesssim
1$~Gyr \citep[e.g.][]{2018ApJ...853..177P,2019MNRAS.490.3025R}. This
suggests that strong thermal instability develops when $t_{\rm
cool}\lesssim 1$~Gyr
\citep{2008ApJ...683L.107C,2008ApJ...687..899R,2009A&A...501..835M,2010A&A...513A..37H,2017MNRAS.464.4360M}. In
this study, we define the threshold radius $r_{\rm i}$ as the one where
the cooling time is $t_{\rm cool}=0.7$~Gyr, which is the typical value
observed for cool-core clusters
\citep{2018ApJ...853..177P,2019MNRAS.490.3025R}. We assume that the cold
interstellar gas is distributed at $r<r_{\rm i}$. On the other hand, the
hot gas loses its thermal energy even at $r>r_{\rm i}$ and we assume
that the cooling is effective up to $r_{\rm cool}=f_{\rm cool}r_{\rm i}$,
where $f_{\rm cool}$ is a constant. Observations have shown that optical
filaments, which are apparently formed through radiative cooling,
surround molecular gas at the center of clusters. The size of the
optical filaments is $\sim 4$ times larger than that of molecular gas
\citep[e.g.][]{2019A&A...631A..22O}. Thus, we adopt $f_{\rm cool}=4$ in
this study.

Previous studies have indicated that the thermal instability develops in
the hot gas when $t_{\rm cool}/t_{\rm ff}\lesssim 20$, where $t_{\rm
ff}=\sqrt{2\: r/g}$ is the free-fall time, and $g$ is the gravitational
acceleration
\citep{2012ApJ...746...94G,2012MNRAS.419.3319M,2015ApJ...811...73L,2015ApJ...811..108P,2015ApJ...799L...1V,2015Natur.519..203V,2017ApJ...845...80V,2016ApJ...830...79M}. In
our fiducial model, $t_{\rm cool}/t_{\rm ff}\sim 20$ at $r=r_{\rm i}\sim
11$~kpc, which means that the thermal instability should develop.

\subsection{Circumnuclear disk and AGN feedback}
\label{sec:CND}

\subsubsection{Disk radius and stability}

The gas of the circumnuclear disk is supplied from the interstellar cold
gas. We assume that the interstellar cold gas that is not
consumed in star formation eventually gathers at the galaxy center and
forms the circumnuclear disk, even if the gas has been warmed up by
the stars. This is because the interstellar cold gas is deposited from
the hot gas that is in hydrostatic equilibrium, which may mean that the
cold gas has little angular momentum. Thus, the cold gas that does not
form stars should almost radially fall toward the galactic center, and
the circumnuclear disk is formed through the residual angular
momentum.

We construct a model of the disk based on those developed by
\citet{2008ApJ...681...73K} and \citet{2013A&A...560A..34W}. For the
sake of simplicity, we do not consider the radial structure of the disk
in contrast with those previous studies. Thus, the disk is represented
as a one-zone system in this study.

For the sake of simplicity, we assume that the disk radius $r_{\rm d}$
is the farthest reach of the dominion of the gravitational potential of
the disk with respect to the host galaxy
\citep{2013A&A...560A..34W}. Thus, it can be derived by numerically
solving equation of
\begin{equation}
\label{eq:MBH}
 M_{\rm BH} + M_{\rm g,d} = M_{\rm host}(r_{\rm d})\:,
\end{equation}
where $M_{\rm BH}$ is the mass of the SMBH and $M_{\rm g,d}$ is the gas
mass of the disk, which is given by solving Equation~(\ref{eq:dMgd}). In
the fiducial model, the disk size is $r_{\rm d}\sim 0.4$~kpc.

The disk has two states: (i) a gravitationally unstable state and (ii)
a gravitationally stable state. We adopt Toomre's stability criterion to
distinguish them (\citealt{1964ApJ...139.1217T}; see also \citealt{2013MNRAS.433.1389R}). The critical surface density is defined as
\begin{equation}
\label{eq:Sigcrit}
 \Sigma_{\rm crit} = \frac{\kappa c_{\rm s,d}}{\pi G}\:,
\end{equation}
where $\kappa^2=4\Omega(r_d)^2 + 2\Omega(r_d)r d\Omega/dr$ is the
epicyclic frequency, and $c_{\rm s,d}$ is the sound velocity of the disk
gas.  We assume that $c_{\rm s,d}=1\rm\: km\: s^{-1}$ following
\citet{2008ApJ...681...73K}. The angular velocity $\Omega(r)$ is given
by
\begin{equation}
\label{eq:Omega}
 \Omega(r)^2 = \frac{G(M_{\rm BH}+M_{\rm host}(r))}{r^3}
 + \frac{\pi G\Sigma_d}{r}\:.
\end{equation}
The surface density of the disk is represented by $\Sigma_{\rm d}=M_{\rm
g,d}/(\pi r_{\rm d}^2)$. The stability criterion indicates that if
$\Sigma_{\rm d}>\Sigma_{\rm crit}$ ($\Sigma_{\rm d}<\Sigma_{\rm crit}$),
the disk is unstable (stable). This corresponds to $Q<1$ (unstable) and
$Q>1$ (stable) in terms of the Toomre $Q$ parameter.

\subsubsection{Unstable disk}

When the disk is unstable ($\Sigma_{\rm d}>\Sigma_{\rm crit}$), star
formation is triggered, and the disk is supported by the turbulent
pressure associated with supernovae (SNe) explosions in the disk. Thus, the
pressure is given by $\rho_{\rm d}v_{\rm t,d}^2$, where $\rho_{\rm d}$
is the disk gas density, and $v_{\rm t,d}$ is the turbulent velocity. The
pressure is balanced with the vertical gravity of the disk $g_{\rm d}$:
\begin{equation}
\label{eq:rhogd}
 \rho_{\rm d}v_{\rm t,d}^2 = \rho_{\rm d}g_{\rm d} h\:,
\end{equation}
where $h$ is the scale height of the disk. The surface density of the
disk is also represented as $\Sigma_{\rm d}=2\:\rho_{\rm d}h$. The
gravity component perpendicular to the disk is
\begin{equation}
\label{eq:g}
 g_{\rm d} = \frac{G(M_{\rm BH}+M_{\rm host}(r_{\rm d})) h}{r_{\rm d}^3}
 + \pi G\Sigma_{\rm d}\:.
\end{equation}

Since the turbulence is driven by SN explosions, the energy
balance can be expressed as
\begin{equation}
\label{eq:diskbalance1}
 \frac{\rho_{\rm d}v_{\rm t,d}^2}{t_{\rm dis,d}}
 = \eta S_* E_{\rm SN}
\end{equation}
where $\eta$ is the heating efficiency, $S_*$ is the star formation rate
per unit volume and time, and $E_{\rm SN} (=10^{51}\rm\: erg)$ is the
energy input by a SN explosion. Since the dissipation time of the
turbulence is given by $t_{\rm dis,d}=h/v_{\rm t,d}$ and the star
formation rate is represented by $S_*=C_*\rho_{\rm d}$, where $C_*$ is
the star formation efficiency, Equation~(\ref{eq:diskbalance1}) can be
rewritten as
\begin{equation}
\label{eq:cd3}
\frac{v_{\rm t,d}^3}{h}
 = \eta C_* E_{\rm SN}
\end{equation}
From Equations~(\ref{eq:rhogd}), (\ref{eq:g}), (\ref{eq:cd3}), one can
obtain
\begin{equation}
 h^{1/2}\left[\frac{G(M_{\rm BH}+M_{\rm host}(r_{\rm d})) h}{r_{\rm d}^3}
 + \pi G\Sigma_{\rm d}\right]^{3/2}=\eta C_* E_{\rm SN}
\end{equation}
The disk height $h$ can be derived by numerically solving this equation,
and then the turbulent velocity $v_{\rm t,d}$ can be obtained from
Equation~(\ref{eq:cd3}). We assume that $\eta=10^{-3}\: M_\odot^{-1}$
and $C_*=3\times 10^{-8}\:\rm yr^{-1}$, following
\citet{2008ApJ...681...73K}. We note that this disk model (see also
\citealt{2020ApJ...889...84K}) is supported by recent observations. For
example, \citet{2021ApJ...914L..11N} detected diffuse synchrotron
emission that is associated with star formation in the circumnuclear
disk of NGC~1275. They also indicated the turbulent velocity of the
circumnuclear disk observed with the Atacama Large
Millimeter/submillimeter Array (ALMA) is consistent with the
model. Moreover, the correlation between the mass of dense molecular gas
around the galactic centers and the SMBH mass accretion rate may also
support the disk model \citep{2016ApJ...827...81I}.

\subsubsection{Stable disk}

When the disk is stable ($\Sigma_{\rm d}<\Sigma_{\rm crit}$), the disk
is supported by thermal pressure $\rho_{\rm d}c_{\rm s,d}^2$, which is
balanced with the vertical gravity of the disk:
\begin{equation}
\label{eq:rhod}
 \rho_{\rm d}c_{\rm s,d}^2 = \rho_{\rm d}g_{\rm d} h\:.
\end{equation}
Thus, the disk height is given by
\begin{equation}
\label{eq:hst}
 h = \frac{c_{\rm s,d}^2}{g_{\rm d}}\:.
\end{equation}
We assume that the star formation in the disk ceases ($C_*=0$) during
the stable state.

\subsubsection{Mass accretion toward SMBH and AGN feedback}
\label{sec:AGN}

We suppose that kinetic viscosity is the cause of angular momentum
transfer in the gas disk. In this case, the mass accretion rate toward
the SMBH is given by
\begin{equation}
\label{eq:dotMBH}
 \dot{M}_{\rm BH} = 2\pi\nu\Sigma_{\rm d}
\left|\frac{d\ln\Omega(r_{\rm d})}{d\ln r_{\rm d}}\right|\:,
\end{equation}
where $\nu$ is the coefficient of kinetic viscosity \citep{1981ARA&A..19..137P}. 

The coefficient $\nu$ depends on the state of the disk. When the disk
is unstable ($\Sigma_{\rm d}>\Sigma_{\rm crit}$), the coefficient is
\begin{equation}
\label{eq:nuv}
 \nu = \alpha v_{\rm t,d} h\:,
\end{equation}
where $\alpha$ is the so-called $\alpha$-parameter. Following
\citet{2008ApJ...681...73K}, we adopt $\alpha=1$, which is based on
numerical simulations \citep{2002ApJ...566L..21W}.  On the other hand,
when the disk is stable ($\Sigma_{\rm d}<\Sigma_{\rm crit}$), the
coefficient is
\begin{equation}
\label{eq:nuc}
 \nu = \alpha c_{\rm s,d} h\:.
\end{equation}
In this case, the magnetorotational instability could be a source of
turbulence. The turbulent velocity is comparable to or even smaller than
the sound velocity, and the $\alpha$-parameter is represented by
$\alpha\sim 0.01$--0.5
\citep[e.g.][]{1991ApJ...376..214B,2000ApJ...532L..67M,2003ApJ...585..429M}.
In this study, we adopt $\alpha=0.05$.

The potential energy of the gas accreted by the SMBH is converted to
radiation, winds or jets, which could heat the galaxy
(AGN feedback). The energy input rate  by the AGN is
\begin{equation}
\label{eq:LAGN}
 L_{\rm AGN} = \epsilon_{\rm heat} \dot{M}_{\rm BH} c^2\:,
\end{equation}
where $\epsilon_{\rm heat}$ is the heating efficiency, and we adopt
$\epsilon_{\rm heat}=0.02$ \citep{2016MNRAS.462.3854L}. We assume that
the AGN feedback acts only on the hot gas, because the volume filling
factor of the interstellar cold gas and circumnuclear gas is very small.
If AGN feedback is effective, it should prevent the hot gas from
cooling. Since the actual feedback mechanism has not been well
understood and it may be quite complicated, such as sound waves
\citep{2006MNRAS.366..417F,2017MNRAS.464L...1F,2018ApJ...858....5Z},
shocks \citep{2015ApJ...805..112R,2017ApJ...847..106L}, cosmic rays
\citep{1991ApJ...377..392L,2008MNRAS.384..251G,2011ApJ...738..182F,2012ApJ...746...53F,2013MNRAS.432.1434F,2013ApJ...779...10P,2017MNRAS.467.1449J,2017ApJ...844...13R,2020MNRAS.491.1190S},
and mixing
\citep{2004ApJ...612L...9F,2016MNRAS.455.2139H,2017MNRAS.466L..39H,2020ApJ...896..104H,2020MNRAS.494.5507F,2020ApJ...892..100U,2021arXiv210607168U},
we represent it with a simple 'switch' for the sake of simplicity. This
means that if the AGN feedback is strong enough, it switches off the
development of thermal instability in the hot gas. If not, the
instability sets in. Thus, we give the suppression factor as
\begin{equation}
\label{eq:fsup}
 f_{\rm sup} = \exp(-L_{\rm AGN}/L_{\rm cool})
\end{equation}
in Equation~(\ref{eq:dMacc}). We assume that the hot gas is
instantaneously heated by the AGN feedback. We have confirmed that the
results are almost the same even if the heating is delayed by the sound
crossing time of the hot gas over the galaxy ($r_{\rm gal}$).

\subsection{Evolution equations for gas and stars}
\label{eq:evo}

Evolution equations for the gas components are constructed based on the
flow chart shown in Figure~\ref{fig:flow}. For the interstellar cold
gas, it is
\begin{equation}
\label{eq:dMgi}
 \dot{M}_{\rm g,i} = \dot{M}_{\rm cool}
 - \frac{M_{\rm g,i}}{t_{\rm dis,c}(M_{\rm g,i})}\:,
\end{equation}
The first term on the right hand side is the supply from the hot gas, and
the second term represents the disruption by star formation. The
evolution of the circumnuclear disk gas is
\begin{equation}
\label{eq:dMgd}
 \dot{M}_{\rm g,d} = \dot{M}_{\rm sup} - \dot{M}_{\rm BH} - C_* M_{\rm g,d}\:.
\end{equation}
where
\begin{equation}
 \label{eq:dMsup}
 \dot{M}_{\rm sup} = (1 - \epsilon_{\rm *,c}(M_{\rm g,i}))\frac{M_{\rm g,i}}{t_{\rm dis,c}(M_{\rm g,i})}\:.
\end{equation}
The first term on the right hand side of Equation~(\ref{eq:dMgd}) or
$\dot{M}_{\rm sup}$ represents the supply from the cold interstellar
gas. The second term is the gas flow toward the SMBH, and the third term
is the consumption by star formation in the disk. 

The mass of the stars formed in the interstellar cold gas $M_{\rm *,i}$
evolves as:
\begin{equation}
\label{eq:dMsi}
 \dot{M}_{\rm *,i} = (1-R_{\rm ret})
\frac{\epsilon_{\rm *,c}(M_{\rm g,i})M_{\rm g,i}}{t_{\rm dis,c}(M_{\rm g,i})}\:,
\end{equation}
where $R_{\rm ret}$ is the fraction of the initial mass of a stellar
population that is returned to the hot gas through mass loss from dying
stars. In the instantaneous recycling approximation, it is $R_{\rm
ret}\sim 0.5$ \citep{2016MNRAS.462.3854L}. The right-hand side of
Equation~(\ref{eq:dMsi}) corresponds to arrow (a) in
Figure~\ref{fig:flow}, although the fraction of $R_{\rm ret}$ goes to
the hot gas (arrow (b)). The evolution of the mass of the stars formed
in the circumnuclear disk $M_{\rm *,d}$ is
\begin{equation}
\label{eq:dMsd}
 \dot{M}_{\rm *,d} = (1-R_{\rm ret}) C_* M_{\rm g,d}\:.
\end{equation}
The right-hand side represents arrows (c) and (d).

Equations~(\ref{eq:dMsi}) and (\ref{eq:dMsd}) show that both $M_{\rm
*,i}$ and $M_{\rm *,d}$ are increasing functions of time. In our
calculations, $M_{\rm *,d}$ far exceeds $M_{\rm g,d}$ over time (see
Figure~\ref{fig:mass}(a))\footnote{This does not happen in previous
studies \citep{2008ApJ...681...73K,2013A&A...560A..34W}, because their
calculation times are shorter than ours.}. However, it is unlikely that
those stars remain in the disk, because a thin heavy stellar disk is
unstable \citep{2008gady.book.....B}. In reality, the stars are likely
to be scattered through the gravitational interaction with the disk
matter when the disk becomes gravitationally unstable even before the
stellar mass dominates in the disk. Thus, we assume that the stars
represented by $M_{\rm *,d}$ are spherically and isothermally
distributed at $r<r_{\rm d}$. This means that the density profile of
those stars is represented by $\propto r^{-2}$ for $r<r_{\rm d}$. Since
the interstellar cold gas is distributed at $r<r_{\rm i}$
(Section~\ref{sec:coldgas}), we also assume that the stars formed in the
cold gas ($M_{\rm *,i}$) are isothermally distributed at $r<r_{\rm
i}$. The potential associated with both $M_{\rm *,i}$ and $M_{\rm *,d}$
is represented by $\Phi_{\rm nstar}(r)$ in
Equation~(\ref{eq:Phi}). However, we note that the the contribution of
$\Phi_{\rm nstar}(r)$ to the whole potential is minor in our
calculations, and it hardly affects the results. For example, the
temperature of the hot gas, which is given by Equation~(\ref{eq:Thot}),
changes only by $\lesssim 0.1$~keV during our calculations.

\begin{deluxetable*}{cccccccc}
\label{tab:average}
\tablecaption{Model Results}
\tablewidth{0pt}
\tablehead{\colhead{Model} & \colhead{$\langle \dot{M}_{\rm BH}\rangle$} & \colhead{$\dot{M}_{\rm cool,0}$} & \colhead{$\langle \dot{M}_{\rm cool}\rangle$} & \colhead{$\langle L_{\rm AGN}\rangle$} & \colhead{$L_{\rm cool}$} & \colhead{$\langle f_{\rm sup}\rangle$} & \colhead{$f_{\rm BH}$}\\
 & \colhead{$(M_\odot\:\rm yr^{-1})$} & \colhead{$(M_\odot\:\rm yr^{-1})$} & \colhead{$(M_\odot\:\rm yr^{-1})$} & \colhead{$(\rm erg\: s^{-1})$} & \colhead{$(\rm erg\: s^{-1})$} &  &  }
\startdata
FD & 1.0 & 100 & 23 & $1.2\times 10^{45}$ & $5.7\times 10^{43}$ & 0.22 & 0.045 \\
LM & 1.0 & 100 & 23 & $1.2\times 10^{45}$ & $5.7\times 10^{43}$ & 0.22 & 0.045 \\
ND & 0.27 & 113 & 0.76 & $3.1\times 10^{44}$ & $6.1\times 10^{43}$ & $6.7\times 10^{-3}$ & 0.36 \\
LE & 1.7 & 1330 & 111 & $1.9\times 10^{45}$ & $7.7\times 10^{44}$ & 0.084 & 0.015 \\
HE & 0.14 & 0.88 & 0.72 & $1.6\times 10^{44}$ & $4.8\times 10^{41}$ & 0.82 & 0.19 
\enddata
\tablecomments{Quantities are averaged over $t=3$--5~Gyr. Since $\dot{M}_{\rm cool,0}$ and $L_{\rm cool}$ are almost time-independent, we do not add $\langle~~\rangle$.}
\end{deluxetable*}

\section{Model behavior}
\label{sec:behav}

Although our model has many parameters (Table~\ref{tab:symbol})
and appears to be complicated, the state of the model galaxy virtually
depends on a small handful of them. From Equations~(\ref{eq:dMacc}),
(\ref{eq:dMacc0}), and (\ref{eq:LAGN}), we obtain
\begin{equation}
 \frac{L_{\rm AGN}}{L_{\rm cool}}
= \frac{2}{5}\frac{\mu m_p c^2\epsilon_{\rm heat}}{kT_{\rm hot}}
\frac{\dot{M}_{\rm BH}}{\dot{M}_{\rm cool}}f_{\rm sup}
\approx 2050\: \frac{\dot{M}_{\rm BH}}{\dot{M}_{\rm
cool}}f_{\rm sup}\:.
\end{equation}
In the last equation, we assume that $\epsilon_{\rm heat}=0.02$ and
$kT_{\rm hot}=2.2$~keV. As is shown below, $L_{\rm AGN}$, $\dot{M}_{\rm
BH}$, $\dot{M}_{\rm cool}$, and $f_{\rm sup}$ often strongly
fluctuate. Thus, we take their time averages, which are represented by
$\langle~~\rangle$:
\begin{equation}
\label{eq:LLav}
 \frac{\langle L_{\rm AGN}\rangle}{L_{\rm cool}}
\approx 2050\: f_{\rm BH}\langle f_{\rm sup}\rangle\:,
\end{equation}
where $f_{\rm BH}=\langle \dot{M}_{\rm BH}\rangle/\langle \dot{M}_{\rm
cool}\rangle$. We note that, while $f_{\rm sup} = \exp(-L_{\rm
AGN}/L_{\rm cool})$ (Equation~(\ref{eq:fsup})), $\langle f_{\rm
sup}\rangle$ is \textit{not} generally equivalent to $\exp(-\langle
L_{\rm AGN}\rangle/L_{\rm cool})$.

The gas flow and the AGN feedback of the model galaxy basically follow
this relation.  The main characteristic of the hot gas is $L_{\rm cool}$
(or equivalently $\dot{M}_{\rm cool,0}$), which is fixed by giving the profile
(Section~\ref{sec:hot}). The fraction $f_{\rm BH}$ depends on the model
of the interstellar cold gas (Section~\ref{sec:ICG}) and that of the
circumnuclear disk (Section~\ref{sec:CND}). The AGN luminosity $\langle
L_{\rm AGN}\rangle$ also depends on the model of the circumnuclear disk.

\begin{figure}
\plotone{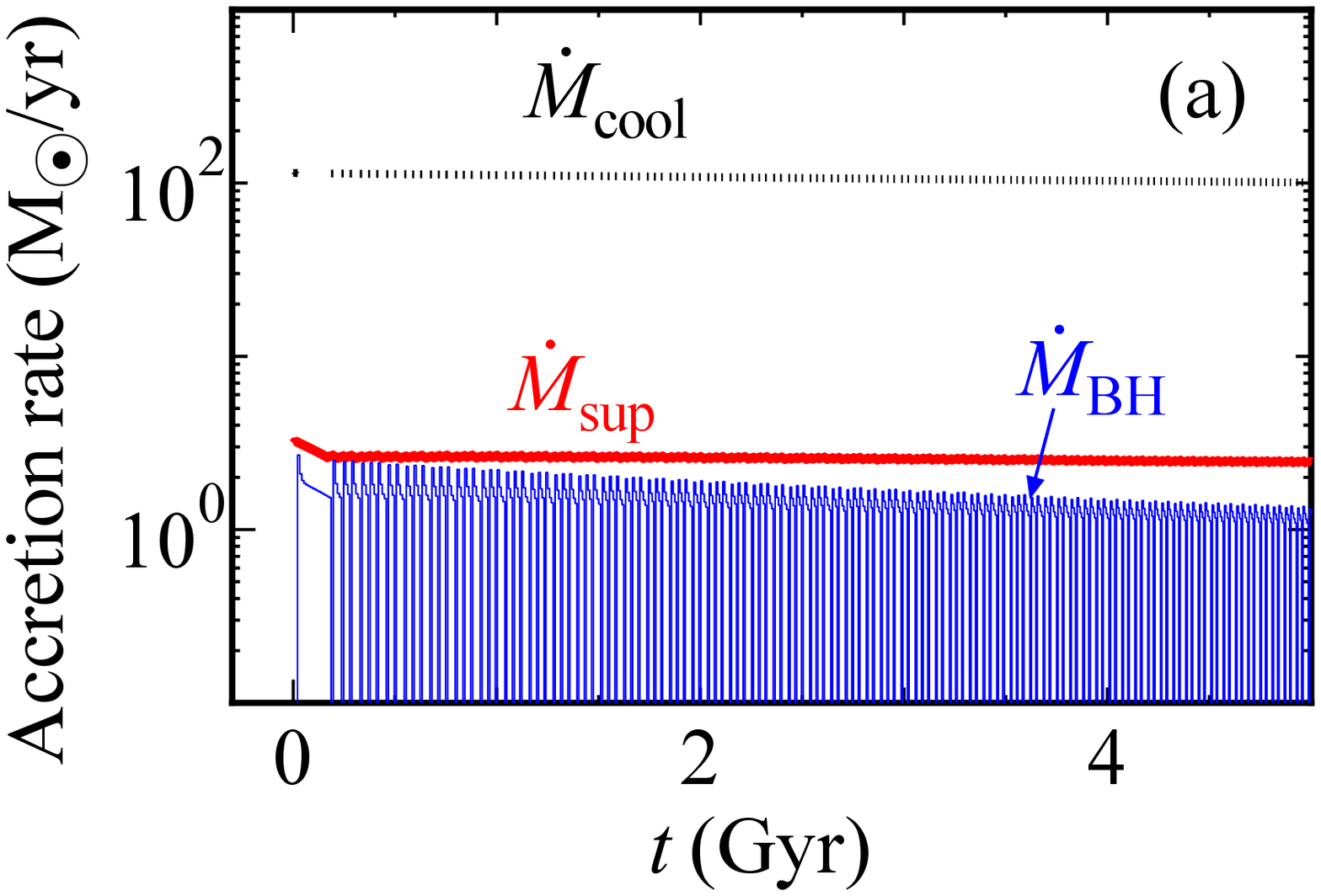} \plotone{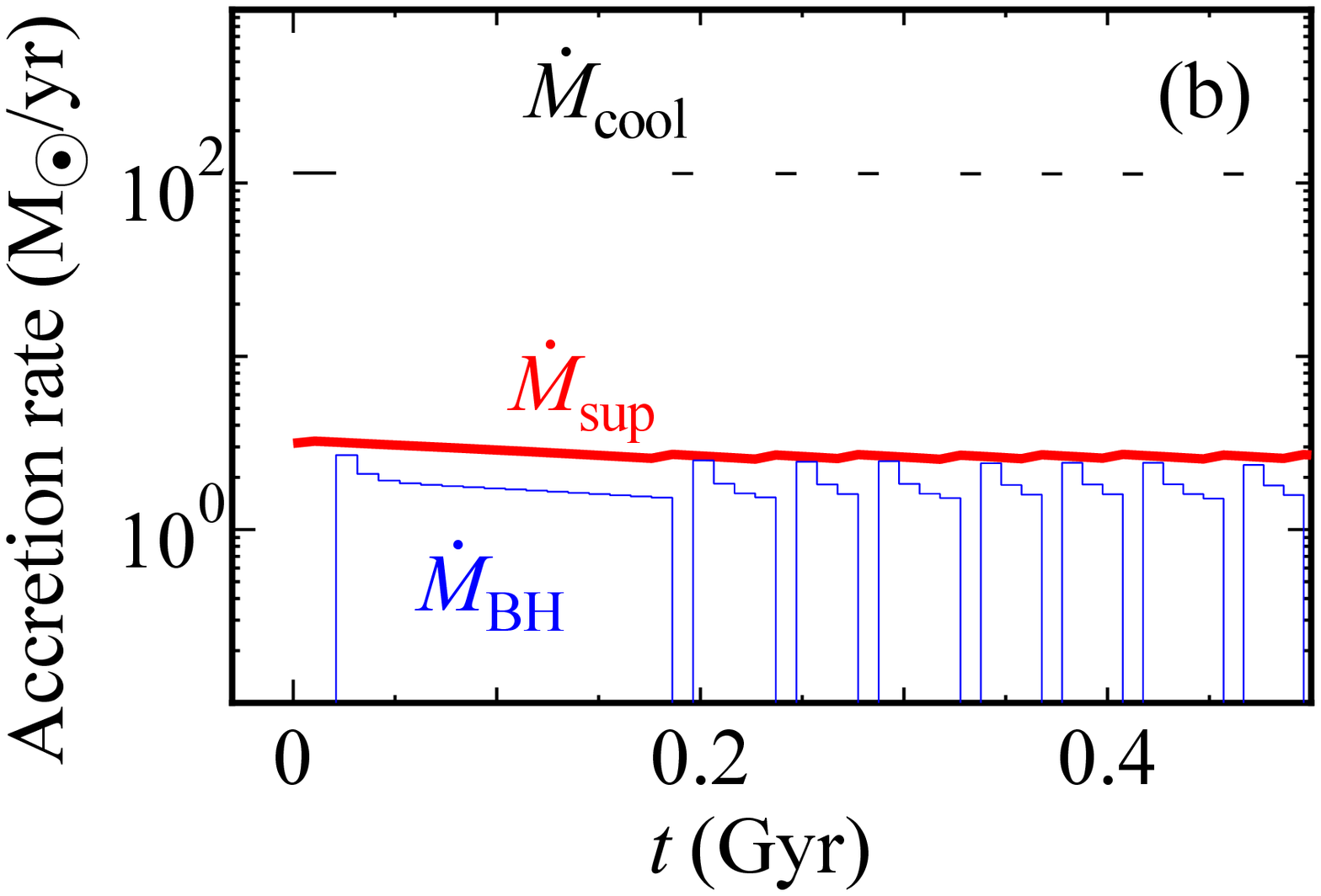} \caption{Evolution of the cooling
rate of the hot gas $\dot{M}_{\rm cool}$ (black dashed line), the supply
rate toward the circumnuclear disk $\dot{M}_{\rm sup}$ (thick red solid
line), and the accretion rate toward the SMBH $\dot{M}_{\rm BH}$ (thin
blue solid line) for the FD model for $t<5$~Gyr (a) and
for $t<0.5$~Gyr (b). Note that the length of each black dash
($\dot{M}_{\rm cool}$) reflects an inactive period of the AGN. The data
of $\dot{M}_{\rm cool}\sim 0$ are not shown for clarity. The mass-supply
rate $\dot{M}_{\rm sup}$ does not show rapid violent changes.}
\label{fig:acc}
\end{figure}

\begin{figure}
\plotone{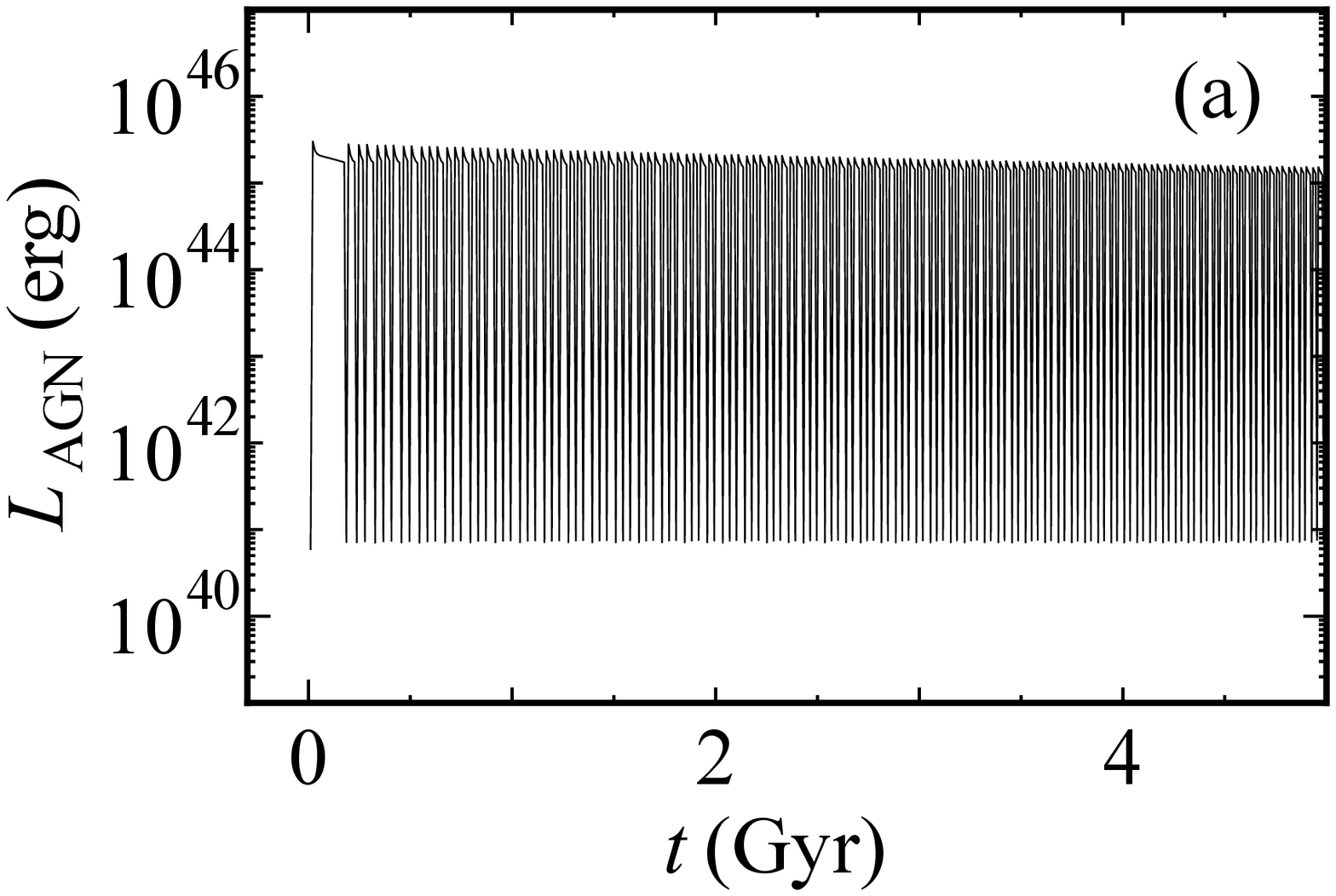}
\plotone{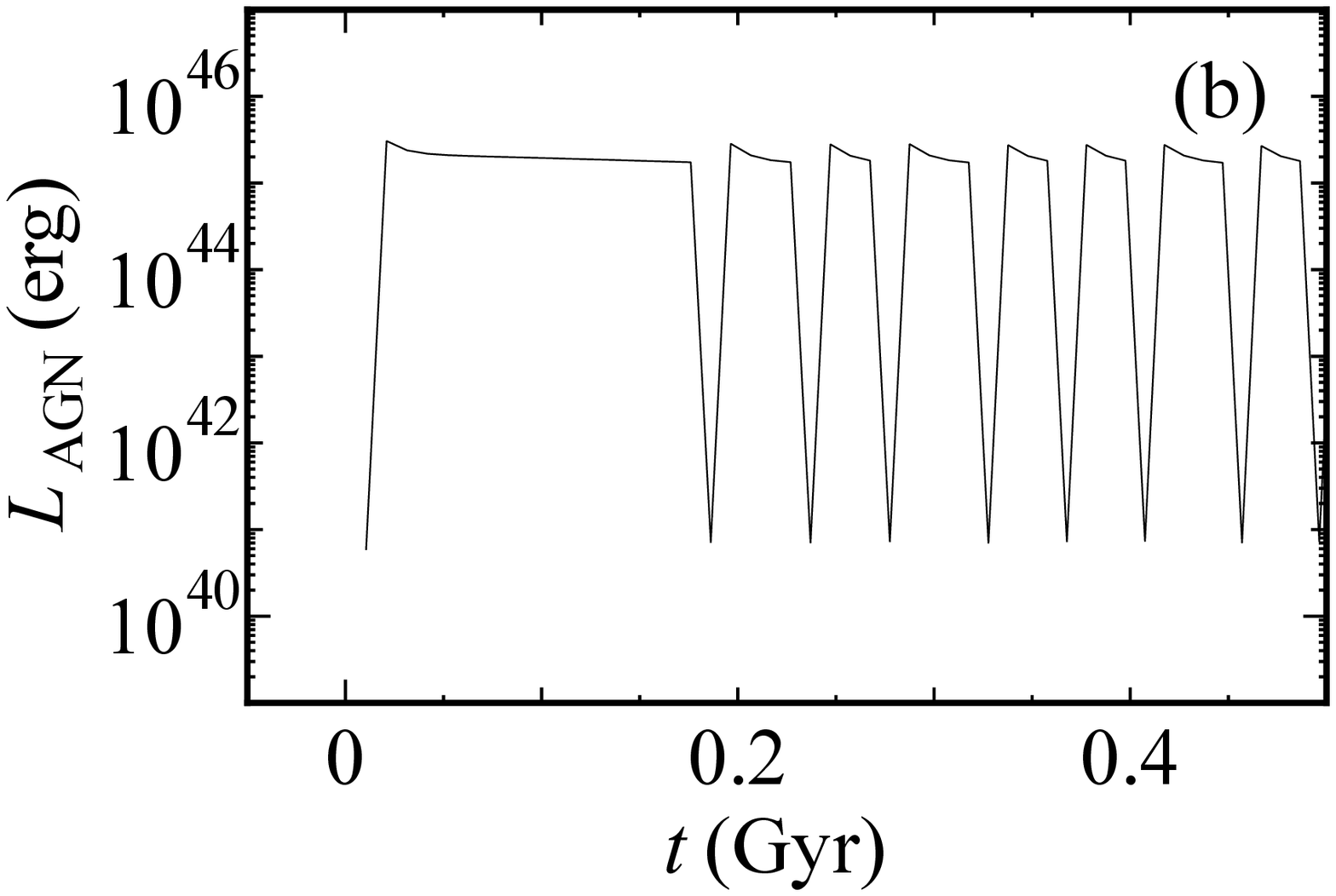}
\caption{Variation
 of AGN power for the FD model for $t<5$~Gyr (a) and for
 $t<0.5$~Gyr (b).}  \label{fig:LAGN}
\end{figure}

\begin{figure}
\plotone{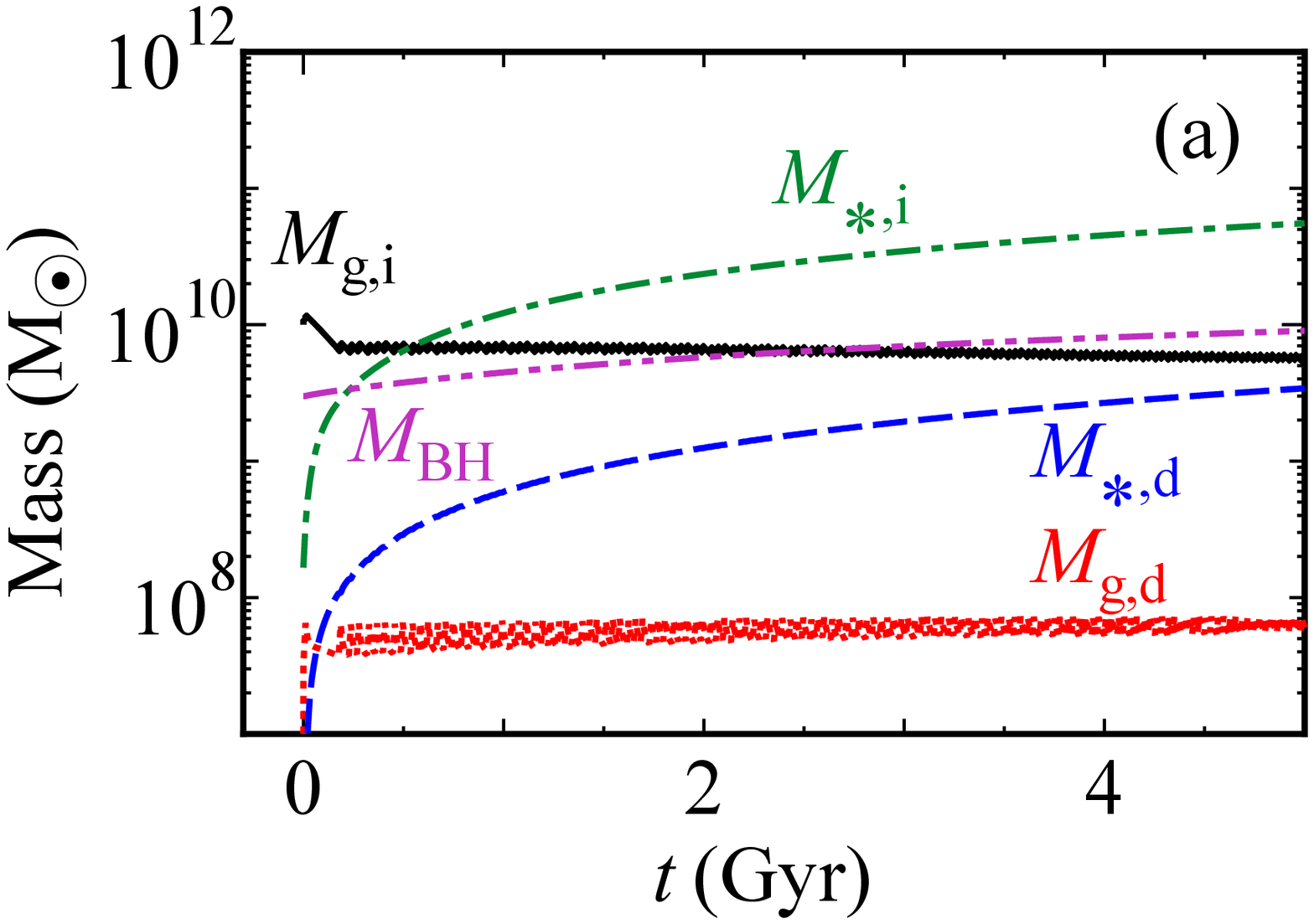} \plotone{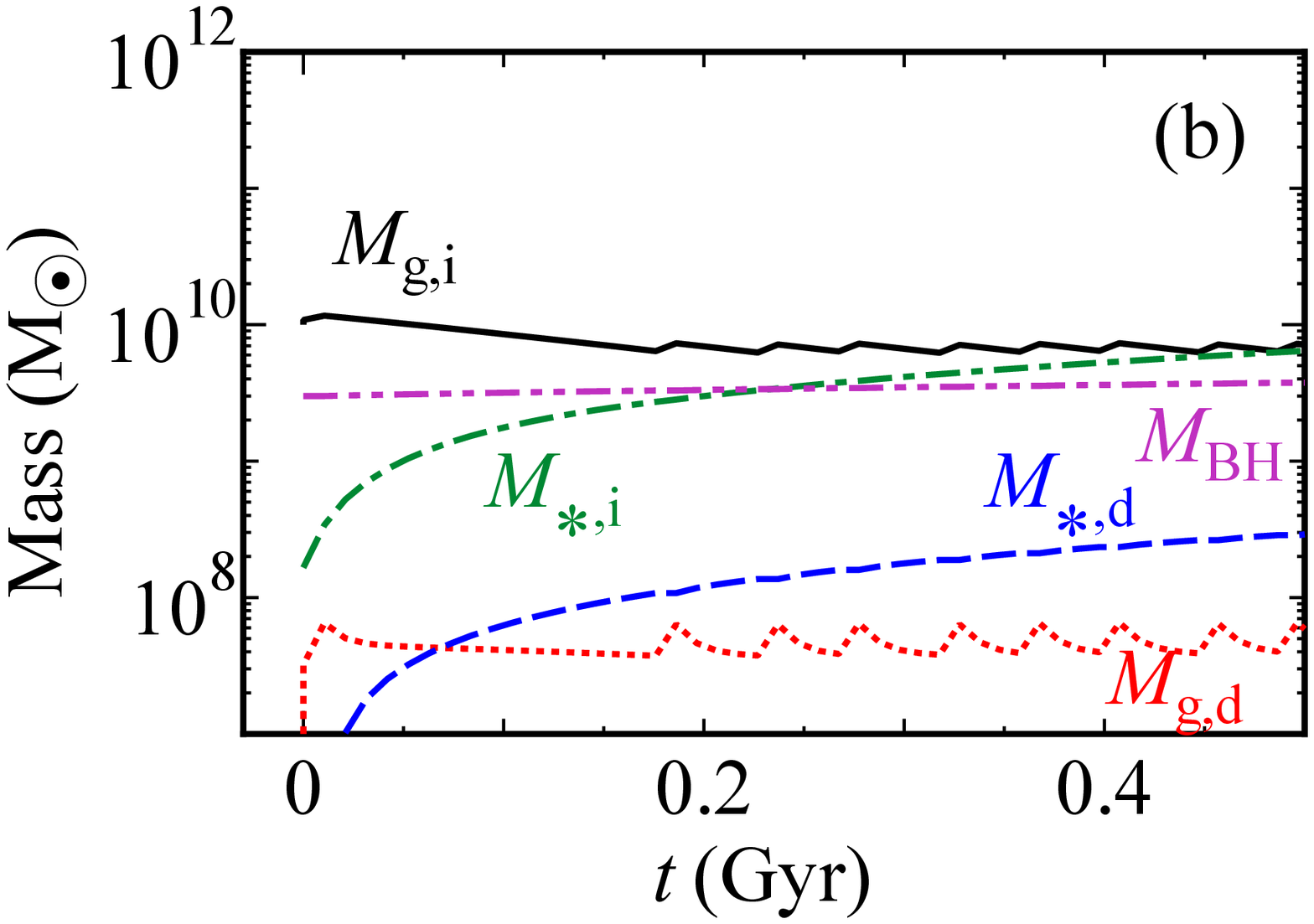} \caption{Evolution of the masses of
the interstellar cold gas $M_{\rm g,i}$ (black solid line), the
circumnuclear disk gas $M_{\rm g,d}$ (red dotted line), the stars formed
in the interstellar cold gas $M_{\rm *,i}$ (blue dashed line), the stars
formed in the circumnuclear disk $M_{\rm *,d}$ (green dashed-dotted
line), and the SMBH $M_{\rm BH}$ (violet dashed-two dotted line) for
model FD for $t<5$~Gyr (a) and for $t<0.5$~Gyr (b).}  \label{fig:mass}
\end{figure}

\begin{figure}
 \plotone{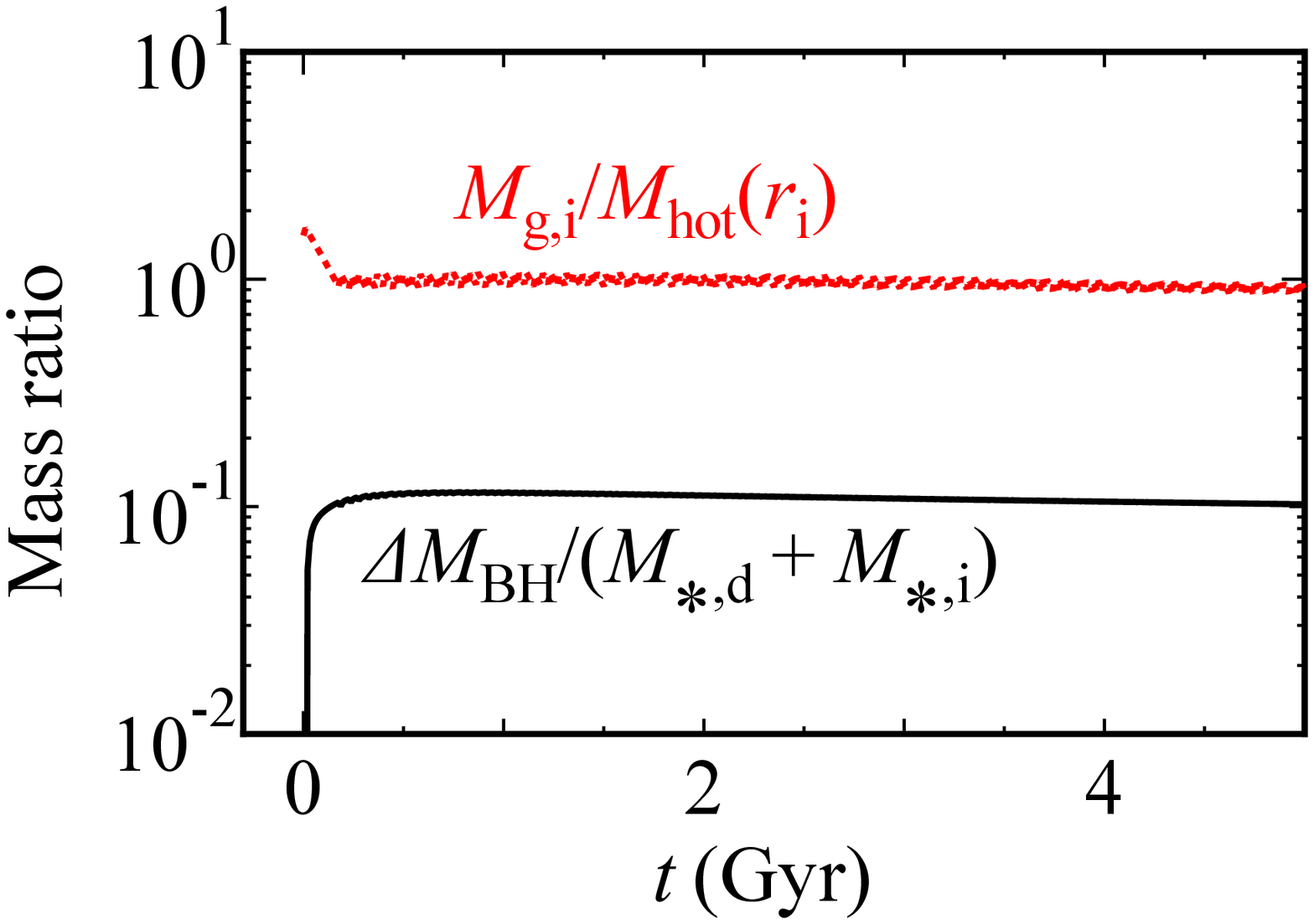} \caption{Black solid line shows the ratio of the
 increment of the black hole mass $\Delta M_{\rm BH}\equiv M_{\rm
 BH}(t)-M_{\rm BH}(0)$ to the mass of newly formed stars $M_{\rm
 *,i}+M_{\rm *,g}$ for model FD. Red dotted line shows the
 ratio of the mass of the cold interstellar gas $M_{\rm g,i}$ to that of
 the hot gas within $r_{\rm i}$, which is represented by $M_{\rm
 hot}(r_{\rm i})$.  } \label{fig:massR}
\end{figure}

\section{Results of the fiducial model}
\label{sec:result}

In the fiducial model (FD model hereafter), we assume that the initial
mass of the SMBH is $M_{\rm BH}=3\times 10^9\: M_\odot$ and that of the
cold interstellar gas is $M_{\rm g,i}=1\times 10^{10}\: M_\odot$. These
are typical values estimated or observed in clusters
\citep[e.g.][]{2011ApJ...727...39M,2019A&A...631A..22O,2019MNRAS.490.3025R}. At
$t=0$, there is no circumnuclear disk ($M_{\rm g,d}=M_{\rm *,d}=0$), and
there are no stars in the cold interstellar gas ($M_{\rm *,i}=0$). We
calculate the evolution of the system by $t=5$~Gyr. We set the time step
of the calculations for $\Delta t=2\pi/\Omega(r_{\rm d})\sim 10^7$~yr,
because our disk model is a one-zone model, and it cannot treat shorter
time-scale phenomena that may happen inside the disk.

Figure~\ref{fig:acc} shows the mass transfer among gas components (see
Figure~\ref{fig:flow}). The mass accretion rate toward the SMBH
($\dot{M}_{\rm BH}$) and the cooling rate of the hot gas ($\dot{M}_{\rm
cool}$) show rapid violent changes at $t\gtrsim 0.2$~Gyr. The variation
of $\dot{M}_{\rm BH}$ reflects the switching between the on and off
states of the gravitational instability of the circumnuclear disk as
follows. If the disk becomes unstable ($\Sigma_{\rm d}> \Sigma_{\rm
crit}$), star formation in the disk is activated. As a result, strong
turbulence develops in the disk, which leads to the efficient transport
of angular momentum and causes an increase in $\dot{M}_{\rm BH}$
(Section~\ref{sec:AGN}). The larger $\dot{M}_{\rm BH}$ causes a decrease
in the mass of the disk gas $M_{\rm g,d}$, which stabilizes the disk
($\Sigma_{\rm d}< \Sigma_{\rm crit}$). During this stable state, the
turbulence in the disk is not strong, and $\dot{M}_{\rm BH}$
decreases. On the other hand, the disk mass increases because of the
stable supply from the interstellar cold gas ($\dot{M}_{\rm sup}$ in
Figure~\ref{fig:acc}), and then the disk becomes unstable again. This
cycle continues to repeat. Since the disk adjusts the AGN fueling so
that the condition of $\Sigma_{\rm d}\sim \Sigma_{\rm crit}$ is kept, we
may call the disk an 'adjusting valve' of AGN feedback.

While a larger $\dot{M}_{\rm BH}$ boosts $L_{\rm AGN}$
(Equation~(\ref{eq:LAGN})), it suppresses $\dot{M}_{\rm cool}$
(Equations~(\ref{eq:dMacc}) and~(\ref{eq:fsup})). Thus, while $L_{\rm
AGN}$ and $\dot{M}_{\rm BH}$ are correlated (Figures~\ref{fig:acc}
and~\ref{fig:LAGN}), $\dot{M}_{\rm cool}$ and $\dot{M}_{\rm BH}$ are
inversely correlated (Figure~\ref{fig:acc}).  In our model, $L_{\rm
AGN}$ just works as a 'switch' of the cooling of the hot gas
(Equation~(\ref{eq:fsup})). Thus, the value of $\epsilon_{\rm heat}$ in
Equation~(\ref{eq:LAGN}) is not important as long as $L_{\rm AGN}\gg
L_{\rm cool}\sim 6\times 10^{43}\rm\: erg\: s^{-1}$
(Table~\ref{tab:average}) or $L_{\rm AGN}\ll L_{\rm cool}$. Even if
$\epsilon_{\rm heat}=0.002$ instead of 0.02 in the FD model, the results
are almost the same except that the AGN luminosity during the active
phase decreases down to $L_{\rm AGN}\sim 10^{44}\rm\: erg\: s^{-1}$.

The gas supply rate $\dot{M}_{\rm sup}$ does not show rapid
sharp changes because it depends on $M_{\rm g,i}$ but not on $\dot{M
}_{\rm cool}$ (Equation~(\ref{eq:dMsup})). This means that the impact of
abrupt changes of the hot-gas cooling rate is alleviated by the
interstellar cold gas, and it does not directly affect the circumnuclear
disk. Thus, the interstellar cold gas works as a 'buffer' that
contributes the stability of AGN feedback.

Figure~\ref{fig:mass} shows that the mass of the interstellar cold gas
$M_{\rm g,i}$ decreases at $t\lesssim 0.2$~Gyr, and then it becomes
almost constant at $t\gtrsim 0.2$~Gyr. During the initial stage
($t\lesssim 0.2$~Gyr), the circumnuclear disk is almost always unstable
because the slightly higher gas-supply rate ($\dot{M}_{\rm sup}$ in
Figure~\ref{fig:acc}) leads to a larger $\Sigma_{\rm d}$.  Thus, the AGN
stays in an active phase ($L_{\rm AGN}\sim 10^{45}\rm\: erg\: s^{-1}$ in
Figure~\ref{fig:LAGN}), and the cooling of the hot gas $\dot{M}_{\rm
cool}$ is strongly suppressed (Equation~(\ref{eq:fsup})). As the mass of
the cold interstellar gas $M_{\rm g,i}$ decreases, $\dot{M}_{\rm sup}$
and $\Sigma_{\rm d}$ also slightly decline. In time, the circumnuclear
disk becomes intermittently stable, and the oscillation of $L_{\rm AGN}$
begins ($t\gtrsim 0.2$~Gyr in Figure~\ref{fig:LAGN}). The final values
of $M_{\rm g,i}\sim 10^{10}\: M_\odot$ and $M_{\rm g,d}\sim 10^8\:
M_\odot$ are consistent with those for NGC~1275 at the center of the
Perseus cluster
\citep{2006A&A...454..437S,2008ApJ...672..252L,2019ApJ...883..193N},
although this may be a result of our choice of parameters
related to the disk stability. We note that the mass of the
interstellar cold gas $M_{\rm g,i}$ is sensitive to the cloud disruption
time $t_{\rm dis,c}$ and the star formation efficiency $\epsilon_{\rm
*,c}$. In the FD model, the disruption time is $t_{\rm
dis,c}\gtrsim 10^8$~yr, which is much longer than that for the giant
molecular clouds in the Milky Way ($t_{\rm dis,c}\sim 10^7$~yr;
\citealt{1997ApJ...480..235E}). This reflects that $t_{\rm dis,c}$ is an
increasing function of the mass $M_{\rm g,i}$, and $M_{\rm g,i}$ of the
model galaxy is much larger than the mass of the giant molecular clouds
in the Milky Way ($\sim 10^5\: M_\odot$; see
Figure~\ref{fig:tdiscM}). Our model shows that the huge masses ($\gtrsim
10^9\: M_\odot$) of observed molecular gas in massive elliptical
galaxies is a consequence of the very long disruption time of the gas,
although the larger $\epsilon_{\rm *,c}$ partially cancels the effect.

Using the time-averaged ($t=3$--5~Gyr) values presented in
Table~\ref{tab:average}, one can show that Equation~(\ref{eq:LLav})
is valid within $\sim 1$\% accuracy. This suggests that the final state
of the galaxy is actually determined by a small number of factors, such
as the cooling rate of the hot gas ($L_{\rm cool}$)\footnote{To be
exact, $L_{\rm cool}$ (and $M_{\rm cool,0}$) slightly decreases ($\sim
10$~\%) during the calculation as the temperature $T_{\rm hot}$ slightly
increases because stars formed in the interstellar cold gas and the
circumnuclear disk deepen the gravitational potential
(Equation~(\ref{eq:Phi})). The increase in $T_{\rm hot}$ leads to a
decrease in $r_{\rm cool}$, which is determined by $t_{\rm cool}\propto
T_{\rm hot}/(\rho(r_{\rm cool})\Lambda(T_{\rm hot},Z))$
(Section~\ref{sec:coldgas}). In Table~\ref{tab:average}, $M_{\rm
cool,0}$ and $L_{\rm cool}$ are also the averaged values although we do
not explicitly show $\langle~~\rangle$.}, the disruption time and the
star formation efficiency of the interstellar cold gas ($t_{\rm dis,c}$
and $\epsilon_{\rm *,c}$), and the stability condition of the
circumnuclear disk that is controlled by the star formation in the disk.

The duty cycle of the AGN activity depends on the time step $\Delta
t$. However, we have confirmed that the overall evolution of the system
and the quantities such as the masses of gas and stellar components are
not sensitive to the time step as long as $\Delta t\lesssim
10^7$~yr. This is because the evolution is mainly determined by the
ratio of the total time when the disk is unstable to that when the disk
is stable. This ratio is not sensitive to $\Delta t$, and it is around 4
for the FD model. We note that, in Figure~\ref{fig:acc}(a), the cooling
rate appears to be $\dot{M}_{\rm cool}\sim 100\: M_\odot\:{\rm
yr^{-1}}\sim \dot{M}_{\rm cool,0}$ (Table~\ref{tab:average}). However,
it actually rapidly changes, which is clearly expressed in a magnified
figure (Figure~\ref{fig:acc}(b)). Since AGN feedback suppresses the
cooling rate down to $\dot{M}_{\rm cool}\sim 0$ for $\sim 80$\% of the
total calculation time, the time-averaged cooling rate is $\langle
\dot{M}_{\rm cool}\rangle\sim 20\: M_\odot\:\rm yr^{-1}$
(Table~\ref{tab:average}). While the evolution of the cold interstellar
gas is determined by $\langle \dot{M}_{\rm cool}\rangle$, it is
insensitive to the period of the rapid change of $\dot{M}_{\rm
cool}$. This may mean that the results do not largely depend on the
details of the AGN feedback as long as the feedback significantly
suppresses the cooling rate $\dot{M}_{\rm cool}$.

Since the stars formed in the galaxy accumulate, the stellar masses
$M_{\rm *,i}$ and $M_{\rm *,d}$ increase as time goes by and exceed the
gas masses $M_{\rm g,i}$ and $M_{\rm g,d}$, respectively
(Figure~\ref{fig:mass}). However, the mass of the newly formed stars,
$M_{\rm *,i}+M_{\rm *,g}$, is much smaller than that of the old stars
that had already existed at $t=0$. In fact, because the old stars are
the origin of the potential $\Phi_{\rm ISO}$
(Equation~(\ref{eq:PhiISO})), their total mass can be calculated from
$\Phi_{\rm ISO}(r_{\rm gal})$, and it is $\sim 3\times 10^{12}\:
M_\odot$.  In Figure~\ref{fig:massR}, we present the ratio of the
increment of the black hole mass $\Delta M_{\rm BH}\equiv M_{\rm
BH}(t)-M_{\rm BH}(0)$ to $M_{\rm *,i}+M_{\rm *,g}$. For $t\gtrsim
0.2$~Gyr, the ratio is almost constant ($\sim 0.1$).
Figure~\ref{fig:massR} also shows the ratio of the mass of the cold
interstellar mass $M_{\rm g,i}$ to that of the hot gas where the cold
gas coexists $M_{\rm hot}(r_{\rm i})$. For $t\gtrsim 0.2$~Gyr, the ratio
is almost constant and the value is close to one, which is consistent
with the results of recent ALMA observations
\citep{2019MNRAS.490.3025R}.

\section{Discussion}
\label{sec:discuss}

\subsection{Roles of the interstellar gas and the circumnuclear disk}
\label{sec:role}

\begin{figure}
 \plotone{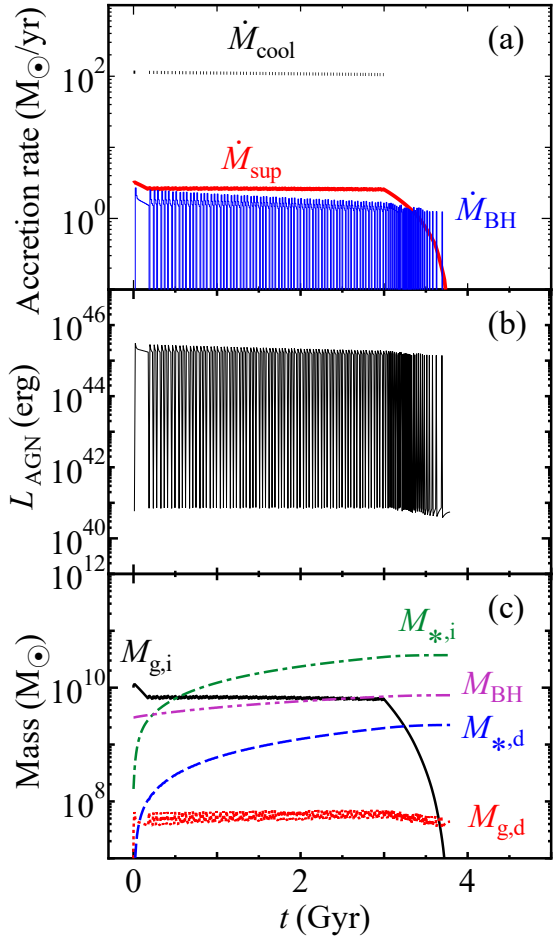} \caption{Results when the cooling of the hot gas
 suddenly stops at $t> 3$~Gyr in model FD. (a) Same as
 Figure~\ref{fig:acc}(a). (b) Same as Figure~\ref{fig:LAGN}(a). (c) Same
 as Figure~\ref{fig:mass}(a).}  \label{fig:stop}
\end{figure}

In order to investigate the role of the cold interstellar gas, we study
the case where the environment of the galaxy suddenly changes. In
Figure~\ref{fig:stop}, we show the results when we intentionally set the
cooling rate of the hot gas at $\dot{M}_{\rm cool}=0$ for $t>3$~Gyr; the
other parameters are the same as those in the FD
model. Figure~\ref{fig:stop}(b) shows that the AGN continues to be
active for $\sim 0.7$~Gyr even after the cooling of the hot gas
stops. This is because the mass supply from the cold interstellar gas to
the nuclear disk continues as long as $M_{\rm g,i}>0$
(Figure~\ref{fig:stop}(c)). The decline time of $M_{\rm g,i}$ is
basically determined by $t_{\rm dis,c}$. This means that the cold
interstellar gas works as a 'fuel tank' for the AGN, which could be
likened to a magma chamber for a volcano. This mechanism may be
important in terms of the stability of AGN feedback. For example, if the
host cluster of the central galaxy undergoes a cluster merger, the hot
gas may suddenly be heated, and its cooling may be halted. Even if this
happens, the fueling of the AGN is not immediately interrupted by this
environmental catastrophe. Thus, the interstellar cold gas may also
work as a 'buffer'.

\begin{figure}
 \plotone{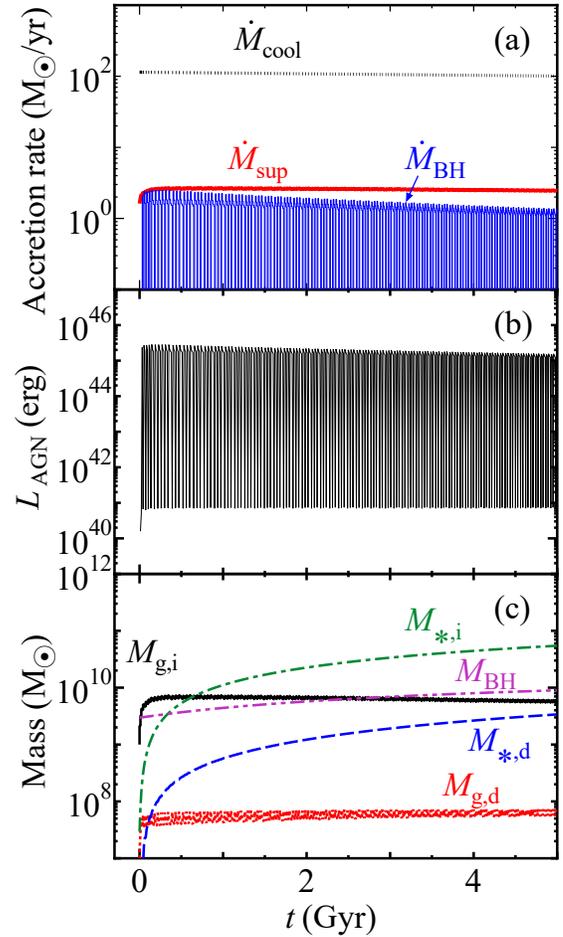} \caption{Results
 when the initial mass of the interstellar cold gas is $M_{\rm
 g,i}=1\times 10^{9}\: M_\odot$ (model LM).  (a) Same as
 Figure~\ref{fig:acc}(a). (b) Same as Figure~\ref{fig:LAGN}(a). (c) Same as
 Figure~\ref{fig:mass}(a).}  \label{fig:lmass}
\end{figure}

The evolution of the system does not depend on the initial condition of
the interstellar cold gas. Figure~\ref{fig:lmass} shows the results when
$M_{\rm g,i}(t=0)=1\times 10^9\: M_\odot$ instead of $1\times 10^{10}\:
M_\odot$ in the FD model. We refer to this low-initial-mass model as model
LM. The mass $M_{\rm g,i}$ rapidly increases for $t<0.2$~Gyr, and then
the evolution of the galaxy is almost the same as that of the FD model
(Figures~\ref{fig:acc}--\ref{fig:mass}). In fact,
Table~\ref{tab:average} shows that parameters related to the mass flows
and powers are the exactly same as those for the FD model. This indicates
that the state of the galaxy is simply described by
Equation~(\ref{eq:LLav}).

\begin{figure}
 \plotone{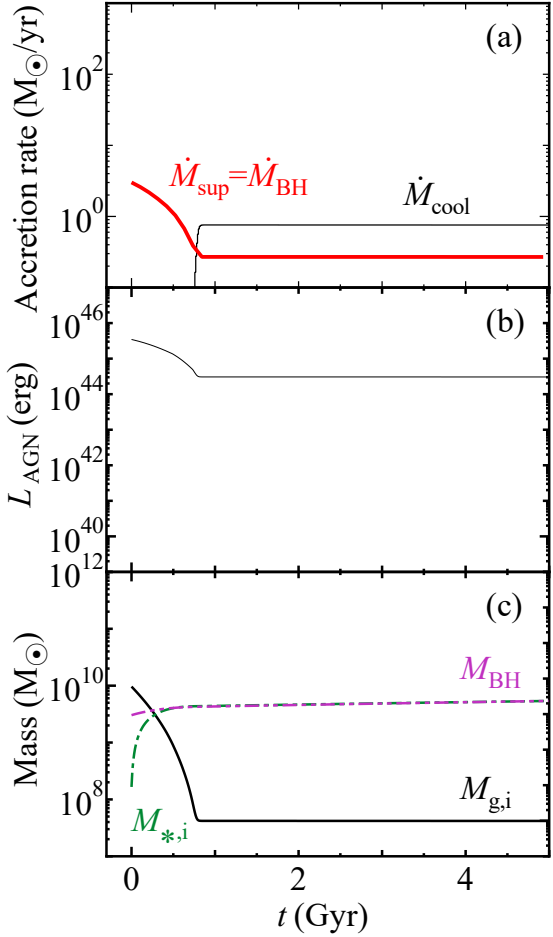} \caption{Results when there is no circumnuclear disk
 (model ND). (a) Same as Figure~\ref{fig:acc}(a) except that
 $\dot{M}_{\rm cool}$ is represented by the thin black solid line. (b)
 Same as Figure~\ref{fig:LAGN}(a). (c) Same as
 Figure~\ref{fig:mass}(a). Since there is no disk, $M_{\rm g,d}$ and
 $M_{\rm *,d}$ are not shown.}  \label{fig:nodisk}
\end{figure}

\begin{figure}
 \plotone{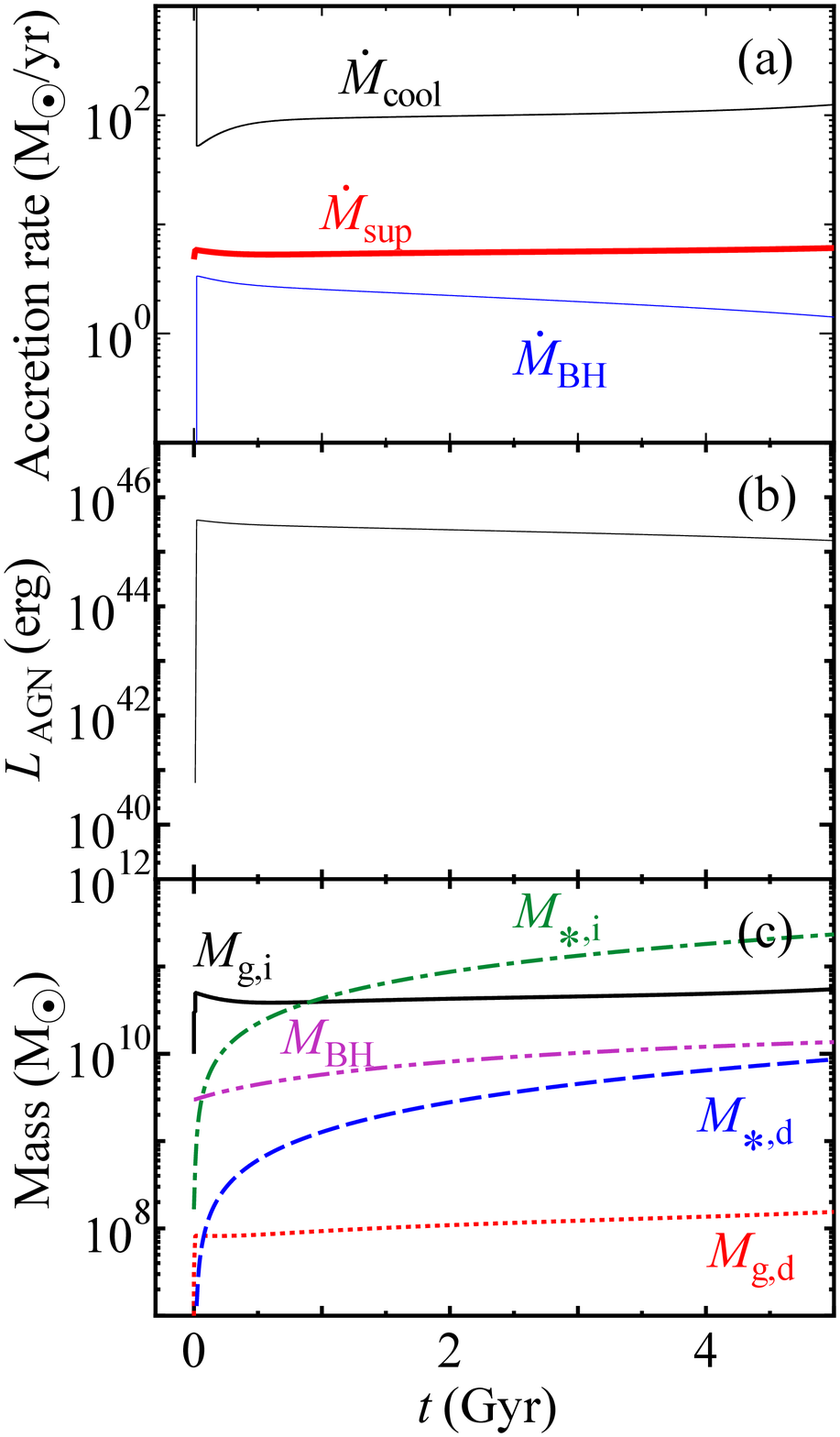} \caption{Results for $n_{\rm e,0}=0.03\rm\: cm^{-3}$,
 which means $K_{30}\sim 22\rm\: keV\: cm^2$ (model LE). (a) Same as
 Figure~\ref{fig:acc}(a) except that $\dot{M}_{\rm cool}$ is represented
 by the thin black solid line. (b) Same as Figure~\ref{fig:LAGN}(a). (c)
 Same as Figure~\ref{fig:mass}(a).}  \label{fig:hne}
\end{figure}

\begin{figure}
 \plotone{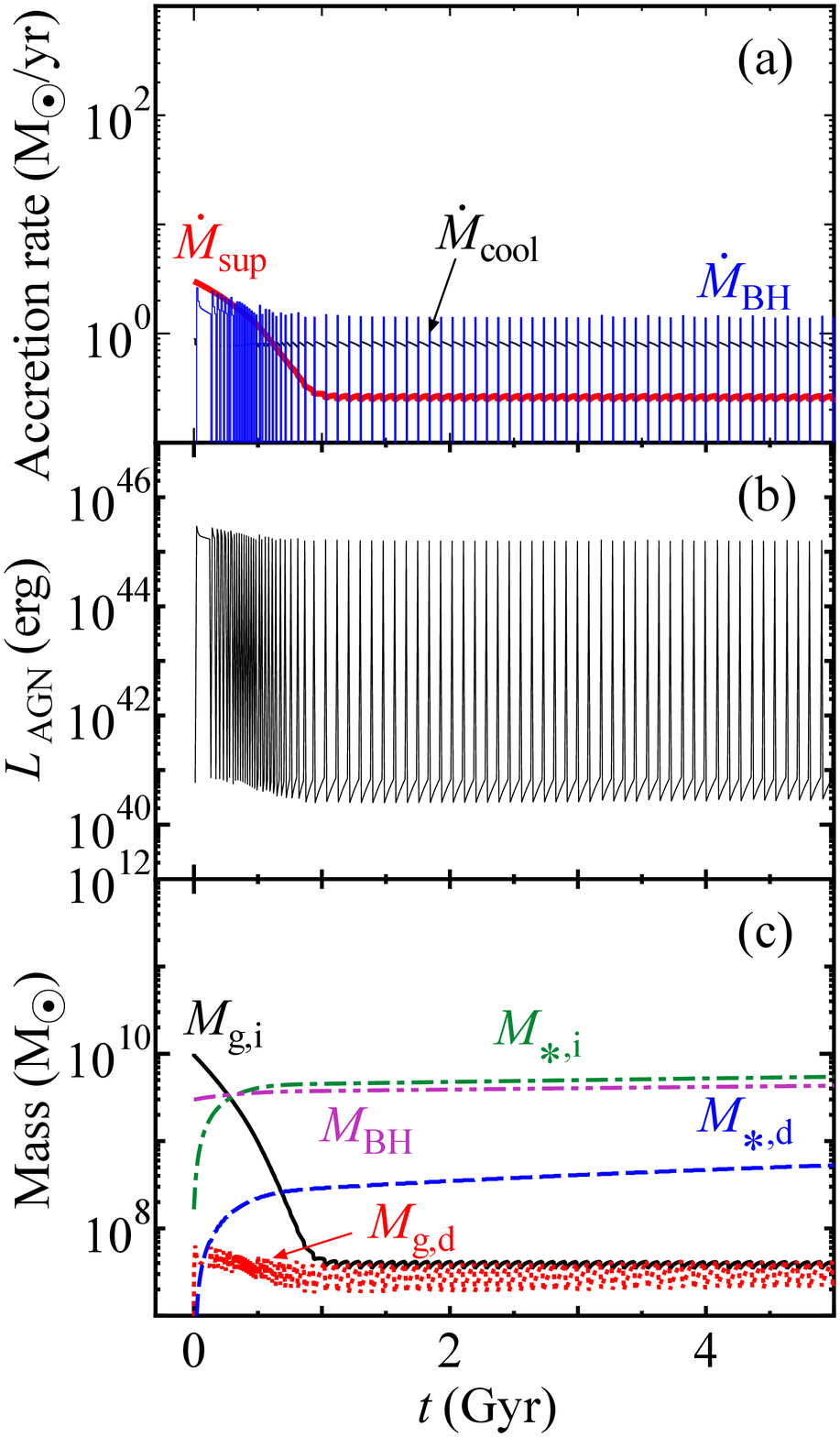} \caption{Results for $n_{\rm e,0}=0.01\rm\: cm^{-3}$,
 which means $K_{30}\sim 46\rm\: keV\: cm^2$ (model HE). (a) Same as
 Figure~\ref{fig:acc}(a) except that $\dot{M}_{\rm cool}$ is represented
 by the thin black solid line. (b) Same as Figure~\ref{fig:LAGN}(a). (c)
 Same as Figure~\ref{fig:mass}(a).}  \label{fig:lne}
\end{figure}

\begin{figure}
 \plotone{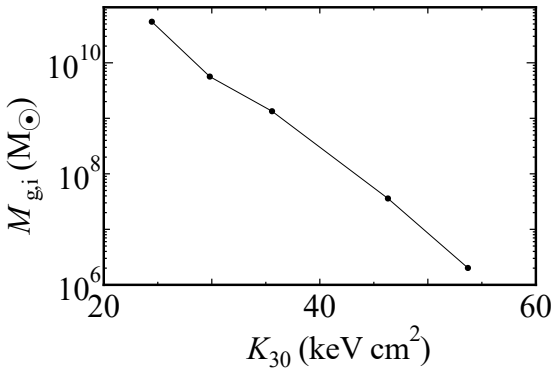} \caption{Relation between the specific entropy at
 $r=30$~kpc ($K_{30}$) and the mass of the interstellar cold gas
 ($M_{\rm g,i}$) at $t=5$~Gyr.}  \label{fig:KM}
\end{figure}

\begin{figure}
 \plotone{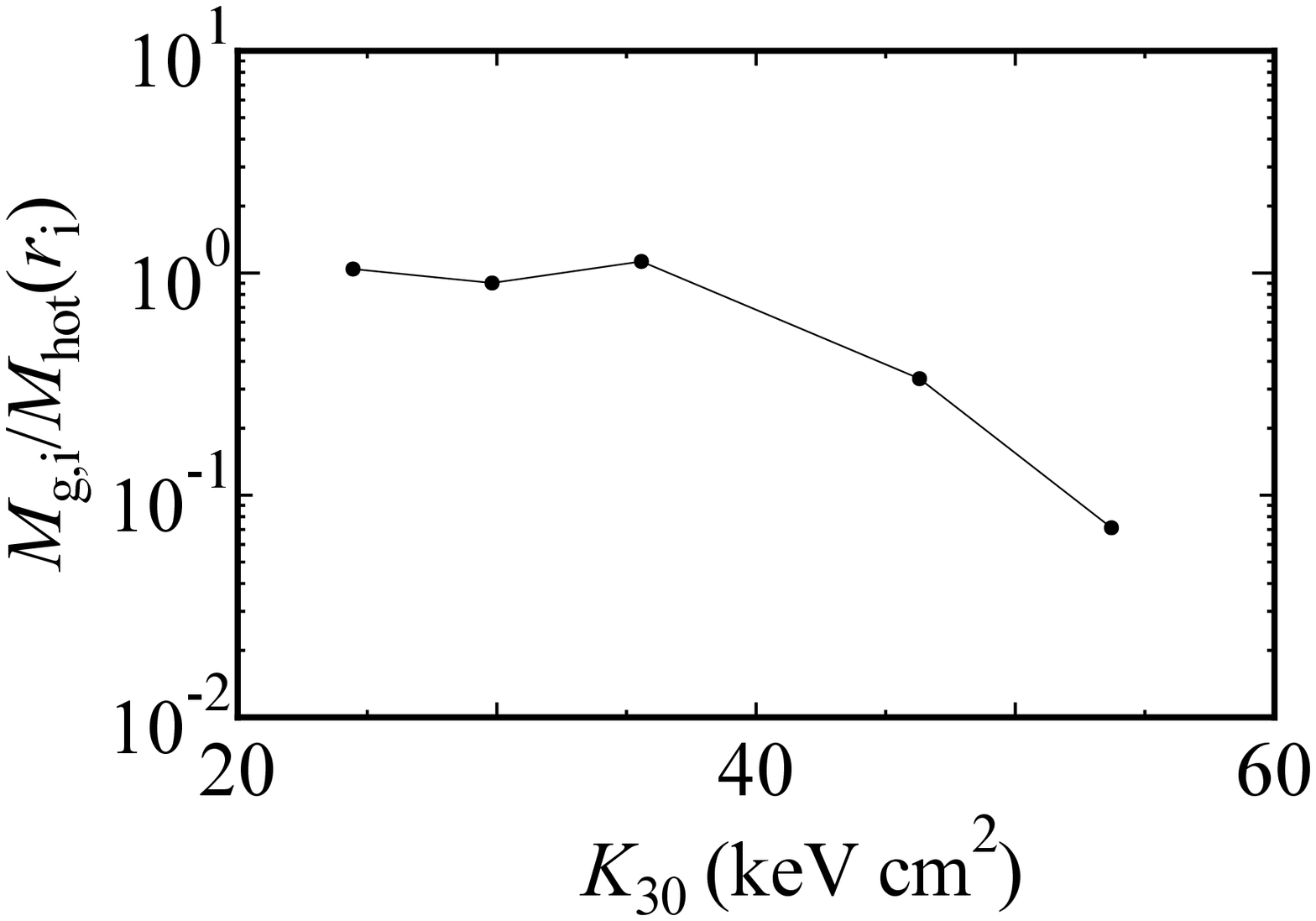} \caption{Relation between the specific entropy at
 $r=30$~kpc ($K_{30}$) and the ratio of the mass of the cold
 interstellar gas ($M_{\rm g,i}$) to that of the hot gas ($M_{\rm
 hot}(r_{\rm i})$) at $t=5$~Gyr.}  \label{fig:MM}
\end{figure}

Figure~\ref{fig:nodisk} shows the results when there is no circumnuclear
disk (model~ND). We consider this situation from a physical point of
view. In this model, the interstellar cold gas directly fuels the SMBH
and we set
\begin{equation}
 \dot{M}_{\rm BH} = \dot{M}_{\rm sup} = (1 - \epsilon_{\rm *,c}(M_{\rm g,i}))\frac{M_{\rm g,i}}{t_{\rm dis,c}(M_{\rm g,i})}\:,
\end{equation}
from Equation~(\ref{eq:dMsup}) instead of Equation~(\ref{eq:dotMBH}). In
contrast with the FD model (Figures~\ref{fig:acc}(a)
and~\ref{fig:LAGN}(a)), the average gas supply rate
($\langle\dot{M}_{\rm BH}\rangle=\langle\dot{M}_{\rm sup}\rangle$) is
low (Table~\ref{tab:average}), and the AGN power $L_{\rm AGN}$ does not
show rapid variation because there is no disk instability. Moreover, the
mass of the interstellar cold gas ($M_{\rm g,i}< 10^8\: M_\odot$;
Figure~\ref{fig:nodisk}(c)) is much smaller than that in the FD model
($M_{\rm g,i}>10^9\: M_\odot$; Figure~\ref{fig:mass}). This is because
the AGN activity {\it constantly} suppresses the cooling of the hot gas,
and the average suppression factor $\langle f_{\rm sup}\rangle$ is much
smaller than that for the FD model (Table~\ref{tab:average}). This
indicates that the presence or absence of the small circumnuclear disk
($r_{\rm disk}\sim 0.4$~kpc) can affect the whole galaxy ($r_{\rm
gal}\sim 30$~kpc) through AGN feedback and can change the amount of the
cold gas in the galaxy. Equation~(\ref{eq:LLav}) is established with a
few percent accuracy for the values in Table~\ref{tab:average}.

In reality, even if the circumnuclear disk is destroyed by a galaxy
collision \citep{2021NatAs...5..478M}, it will soon be rebuilt if there
is interstellar cold gas.  For example, the disk mass $M_{\rm g,d}$
grows rapidly from $M_{\rm g,d}=0$ at $t\sim 0$, as is shown in
Figure~\ref{fig:mass}(b). However, if the cold gas has no angular
momentum and the disk is not formed, the situation similar to
Figure~\ref{fig:nodisk} may be realized.

\subsection{Entropy of the hot gas and the interstellar cold gas}
\label{sec:entro}

Observations have indicated that cold gas (molecular gas) and warm gas
(H$\alpha$-emitting gas) tend to be discovered in galaxies with
low-entropy hot gas
\citep{2008ApJ...683L.107C,2008ApJ...687..899R,2009MNRAS.398.1698S,2017MNRAS.464.4360M}. This
implies a strong connection between the entropy and the formation of the
interstellar cold gas. Motivated by these observations, we study the
connection using our semianalytical model.

Here, we define the specific entropy as $K\equiv kT n_{\rm e}^{-2/3}$
and refer to the value at $r=r_{\rm gal}=30$~kpc as
$K_{30}$. \citet{2017ApJ...851...66H} showed that $K_{30}\sim
25$--$60\rm\: keV\: cm^2$ for clusters with H$\alpha$ nebular emission
and $K_{30}\gtrsim 60\rm\: keV\: cm^2$ for those without H$\alpha$
nebular emission. We note that the entropies at the centers of the
former are $K\lesssim 30\rm\: keV\: cm^2$ \citep{2017ApJ...851...66H}.
Thus, the condition of $K_{30}\sim 25$--$60\rm\: keV\: cm^2$ is roughly
equivalent to the known condition of gas cooling ($K\lesssim 30\rm\:
keV\: cm^2$ at $r\sim 0$;
\citealt{2008ApJ...683L.107C,2008ApJ...687..899R,2009MNRAS.398.1698S}).

In the FD model, $K_{30}\sim 30\rm\: keV\: cm^2$ and the value of
$n_{\rm e}$ at $r=r_{\rm gal}=30$~kpc, namely $n_{\rm e,0}$, is
$0.02\rm\: cm^{-3}$ (Equation~(\ref{eq:rhohot})). In this subsection, we
change the value of $n_{\rm e,0}$, while other parameters are the same
as those for the FD model.  Figure~\ref{fig:hne} shows the evolution
when $n_{\rm e,0}=0.03\rm\: cm^{-3}$, which means $K_{30}\sim 24\rm\:
keV\: cm^2$. We refer to this low-entropy model as model LE.  In this
case, the intrinsic radiative cooling rate of the hot gas ($L_{\rm
cool}$ or equivalently $M_{\rm 0,cool}$) is much larger than that for
the FD model (Table~\ref{tab:average}). Although the constantly large
AGN power $\langle L_{\rm AGN}\rangle$ suppresses the actual gas
cooling, the average gas cooling rate $\langle \dot{M}_{\rm
cool}\rangle$ is still larger than that for the FD model
(Table~\ref{tab:average}).  Thus, the mass of the interstellar cold gas
keeps a large value ($M_{\rm g,i}\sim 5\times 10^{10}\: M_\odot$;
Figure~\ref{fig:hne}) in comparison with that for the FD model ($M_{\rm
g,i}\sim 6\times 10^{9}\: M_\odot$ for $t\gtrsim 1$~Gyr;
Figure~\ref{fig:mass}). As a result, the mass flow rate toward the
circumnuclear disk is high, which means that the disk is constantly
heavy and unstable. Thus, the disk no longer works as the adjusting
valve of mass accretion toward the SMBH. The resultant large
$\dot{M}_{\rm BH}$ supports the large $L_{\rm AGN}$. We have confirmed
that Equation~(\ref{eq:LLav}) is approximately ($\sim 5$~\%) valid for
the values in Table~\ref{tab:average}. In our model, the cold gas that
is not swallowed by the SMBH is consumed in star formation. The star
formation rate of the galaxy is given by
\begin{equation}
\label{eq:dMs}
 \dot{M}_* = \frac{\epsilon_{\rm *,c}(M_{\rm g,i})
M_{\rm g,i}}{t_{\rm dis,c}(M_{\rm g,i})} + C_* M_{\rm g,d}\:,
\end{equation}
and it is relatively high ($\dot{M}_*\sim 100\: M_\odot\:\rm yr^{-1}$)
compared with the average value for the FD model ($\dot{M}_*\sim
20\: M_\odot\:\rm yr^{-1}$ for $t\gtrsim 1$~Gyr). The results are at
least qualitatively similar to what is happening in the Phoenix
cluster. This cluster has an exceptionally low entropy ($K_{30}\lesssim
20\rm\: keV\: cm^2$; \citealt{2019ApJ...885...63M,2020PASJ...72...33K}),
a high star formation rate ($\dot{M}_*\sim 500$--$600\: M_\odot\:\rm
yr^{-1}$; \citealt{2015ApJ...811..111M,2017MNRAS.465.3143M}), and an
active AGN ($L_{\rm AGN}\sim 10^{46}\rm\: erg\: s^{-1}$;
\citealt{2019ApJ...885...63M,2020PASJ...72...62A}). Since
Figure~\ref{fig:hne} shows that the system is stable, the Phoenix
cluster may keep its unusual state unless its environment (e.g. the hot
gas) changes.

In Figure~\ref{fig:lne}, we present the results when $n_{\rm
e,0}=0.01\rm\: cm^{-3}$, which means $K_{30}\sim 46\rm\: keV\: cm^2$.
We refer to this high-entropy model as model HE. The low density and the
high entropy mean that the cooling rate of the hot gas is intrinsically
low ($L_{\rm cool}$ or $M_{\rm cool,0}$ in Table~\ref{tab:average}). The
actual cooling rate of the hot gas $\dot{M}_{\rm cool}$
(Figure~\ref{fig:lne}(a)) is also low compared with that of the FD model
(Figure~\ref{fig:acc}(a) and Table~\ref{tab:average}). As a result, the
mass of the interstellar cold gas $M_{\rm g,i}$ decreases at $t\lesssim
1$~Gyr; the decline time of $M_{\rm g,i}$ is mainly determined by
$t_{\rm dis,c}$. Then $M_{\rm g,i}$ maintains a small value ($M_{\rm
g,i}\sim 4\times 10^7\: M_\odot$; Figure~\ref{fig:lne}(c)). Since the
gas supply to the circumnuclear disk ($\dot{M}_{\rm sup}$) is also
reduced at $t\gtrsim 1$~Gyr (Figure~\ref{fig:lne}(a)), the disk gas mass
$M_{\rm g,d}$ (Figure~\ref{fig:lne}(c)) and the surface density
$\Sigma_{\rm d}$ are kept low. Thus, the disk is stable most of the time,
and the activity of the AGN is generally low ($L_{\rm AGN}\sim
2$--$6\times 10^{40}\rm\: erg\: s^{-1}$; Figure~\ref{fig:lne}(b)),
although the AGN often shows short bursts ($L_{\rm AGN}\sim10^{45}\rm\:
erg\: s^{-1}$) because of instantaneous disk instabilities.  The star
formation rate at $t\gtrsim 1$~Gyr is also very small
($\dot{M}_*\lesssim 1\: M_\odot\:\rm yr^{-1}$). Again, we have confirmed
that the final state of the system is represented by
Equation~(\ref{eq:LLav}) with a few percent accuracy using the values
in Table~\ref{tab:average}.

We also study the evolution when the entropy is even higher ($n_{\rm
e,0}=0.006\rm\: cm^{-3}$ and $K_{30}\sim 65\rm\: keV\: cm^2$). In this
case, the interstellar cold gas disappears ($M_{\rm g,i}\rightarrow 0$)
by $t\sim 1$~Gyr, and the AGN activity completely cease. The Virgo
cluster (M87) may be close to this situation \citep{2018MNRAS.475.3004S}
because of the high entropy of $K_{30}\sim 70\rm\: keV\: cm^2$
\citep{2002A&A...386...77M}.

In Figure~\ref{fig:KM}, we summarize the relation between $K_{30}$ and
the mass of the interstellar cold gas $M_{\rm g,i}$ at
$t=5$~Gyr. Figure~\ref{fig:MM} shows the relation between $K_{30}$ and
the ratio of the mass of the cold interstellar gas, $M_{\rm g,i}$, to
that of the hot gas, $M_{\rm hot}(r_{\rm i})$, at $t=5$~Gyr. The
evolution of the ratio was shown in Figure~\ref{fig:massR} in the case
of the FD model ($K_{30}\sim 30\rm\: keV\: cm^2$). In Figure~\ref{fig:MM},
while the ratio is $\sim 1$ for $K_{30}\lesssim 40\rm\: keV\: cm^2$, it
decreases for $K_{30}\gtrsim 40\rm\: keV\: cm^2$. This trend could be
discussed in future observations.

The results in this subsection clearly indicate that the evolution of
the interstellar cold gas and AGN feedback induced by the cold gas are
very sensitive to the entropy of the hot gas. 
Equation~(\ref{eq:LLav}) suggests that $L_{\rm cool}$ is a fundamental
factor that determines the state of the system, and that the entropy is
one manifestation of that because the entropy is anticorrelated with
$L_{\rm cool}$. The entropy we studied ($K_{30}\sim 30\rm\: keV\:
cm^2$) corresponds to our threshold cooling time $t_{\rm cool}=0.7$~Gyr
(Section~\ref{sec:coldgas}), which is comparable to the actual survival
time of the interstellar cold gas ($\sim 0.7$~Gyr;
Section~\ref{sec:role} and Figure~\ref{fig:stop}). This may mean that
the threshold cooling time reflects the condition that the cooling of
the hot gas compensates the dissipation of the interstellar cold gas.
Since the profile of the hot gas is based on the boundary conditions at
$r_{\rm gal}=30$~kpc (Section~\ref{sec:hot}), our results may indicate
that the surrounding intracluster gas of the host cluster 
regulates the formation of the interstellar cold gas and AGN
feedback.  Since cluster mergers can significantly change the
distribution of the intracluster gas, they can affect the mass of the
interstellar cold gas. Moreover, if an enormously powerful AGN explosion
boosts the entropy beyond the boundary ($r>r_{\rm gal}$), it may have a
similar effect on the interstellar cold gas.

\section{Conclusion}
\label{sec:conc}

We have studied the role of massive molecular gas clouds in AGN feedback
in giant elliptical galaxies at the centers of galaxy clusters using a
semianalytical model. We constructed the model based on representative
models for the evolution of the massive molecular gas (interstellar cold
gas) and the circumnuclear disk. We consider the destruction of the
interstellar cold gas by star formation, and we also take into account
the gravitational instabilities of the circumnuclear disk. Our model
reproduced the basic properties of the interstellar cold gas and the
circumnuclear disk, such as their masses. We found that the final state
of a galaxy is represented by a simple relation
(Equation~(\ref{eq:LLav})), which reflects three key factors: (1) the
stability condition of the circumnuclear disk, (2) the disruption time
and the star formation efficiency of the interstellar cold gas, and (3)
the cooling rate or the X-ray luminosity of the hot gas ($L_{\rm
cool}$). The role of each factor can be summarized as follows.

\begin{enumerate}
\item
The
circumnuclear disk tends to stay at the boundary between stable and
unstable states. This works as an 'adjusting valve' and regulates mass
accretion toward the SMBH and AGN feedback. 

 \item
The long disruption time of the interstellar cold gas is the reason for
       its large mass.
The interstellar cold gas
serves as a 'fuel tank' in AGN feedback. Even if the cooling of the
galactic hot gas is halted for some reason (e.g. cluster mergers), the
interstellar cold gas can supply its gas to the circumnuclear disk and
maintain AGN activity for $\gtrsim 0.5$~Gyr. 

\item 
The luminosity $L_{\rm cool}$ regulates the mass supply to the
interstellar cold gas; larger $L_{\rm cool}$ means a larger supply
rate. Since the entropy and the cooling time of the hot gas tend to be
smaller for clusters with larger $L_{\rm cool}$, the mass of the
interstellar cold gas increases as the entropy or equivalently the
cooling time of the hot gas decreases. We confirmed that the small
entropy of the hot gas at the cluster centers ($\lesssim 30\rm\: keV\:
cm^2$ at $r\sim 0$) or the short cooling time ($\lesssim
1$~Gyr) is a key condition for the existence of the massive molecular
gas clouds in the central galaxy, as previous studies suggested. We found that
       if the entropy is
much smaller than $30\rm\: keV\: cm^2$, the star formation and the AGN
become very active, which has actually been observed in the Phoenix
cluster. The ratio of the mass of the interstellar cold gas to that of
the hot gas is close to one when the entropy of the hot gas is
relatively small. The critical cooling time of the hot gas ($\lesssim
1$~Gyr) may be related to the long disruption time
       of the interstellar cold gas.
\end{enumerate}

Since our model is very simple, particularly for the hot gas and AGN
feedback, more sophisticated models and/or numerical simulations would
be desirable. For example, semianalytical models that include the
evolution of hot gas profiles and numerical simulations that resolve
circumnuclear disks would be useful. 


\begin{acknowledgments}
We would like to thank the anonymous referee for a constructive
report. This work was supported by JSPS KAKENHI No.18K03647, 20H00181
(Y.F.), 19K03918 (N.K.), and JP18K03709 (H.N.)
\end{acknowledgments}

\bibliography{cold}{}
\bibliographystyle{aasjournal}

\end{document}